%% file: sample-acmsmall.tex
\documentclass[acmsmall]{acmart}
\usepackage[utf8]{inputenc}
\usepackage[english]{babel}
\usepackage{amssymb}
\usepackage{amsmath}
\usepackage{latexsym}
\usepackage{amsfonts}
\usepackage{wrapfig} % Images NEXT to the Text
\usepackage{graphicx}
\usepackage{rotating}

\usepackage{tikz}
\usetikzlibrary{arrows,shapes,automata,backgrounds,petri,matrix,calc,patterns,fit}

\usepackage{tablefootnote}
\usepackage{arydshln}
\DeclareGraphicsExtensions{.pdf,.jpeg,.png}
\usepackage{graphicx}
\usepackage{caption}
\usepackage{subcaption}
\usepackage{enumerate}
\usepackage{float}
\usepackage{color,soul}
\usepackage{array,colortbl,xcolor}
\usetikzlibrary{tikzmark}
\usetikzlibrary{decorations.pathmorphing,shapes}

\usepackage[framemethod=tikz]{mdframed}
\usepackage{pdflscape}
\usepackage{todonotes}
\usepackage{algorithm}% http://ctan.org/pkg/algorithm
\usepackage{algpseudocode}% http://ctan.org/pkg/algorithmicx
\usepackage{adjustbox}
\usepackage{wrapfig,graphicx,lipsum}

\newcounter{sarrow}

\def\BibTeX{{\rm B\kern-.05em{\sc i\kern-.025em b}\kern-.08emT\kern-.1667em\lower.7ex\hbox{E}\kern-.125emX}}
    
\copyrightyear{2019}
\acmYear{2019}
\setcopyright{acmlicensed}
\acmConference[Woodstock '18]{Woodstock '18: ACM Symposium on Neural Gaze Detection}{June 03--05, 2018}{Woodstock, NY}
\acmBooktitle{Woodstock '18: ACM Symposium on Neural Gaze Detection, June 03--05, 2018, Woodstock, NY}
\acmPrice{15.00}
\acmDOI{10.1145/1122445.1122456}
\acmISBN{978-1-4503-9999-9/18/06}

\begin{document}

%
% The "title" command has an optional parameter, allowing the author to define a "short title" to be used in page headers.
\title{Business Process Variant Analysis: Survey and Classification}

\author{FARBOD TAYMOURI}
\affiliation{%
 \institution{The University of Melbourne}
 \streetaddress{}
 \city{Melbourne}
 \state{}
 \country{AUSTRALIA}}
 
\author{MARCELLO LA ROSA}
\affiliation{%
  \institution{The University of Melbourne}
  \streetaddress{}
  \city{Melbourne}
  \state{}
  \country{AUSTRALIA}}
  
\author{MARLON DUMAS}
\affiliation{%
  \institution{University of Tartu}
  \streetaddress{}
  \city{Tartu}
  \state{}
  \country{ESTONIA}}
  
\author{FABRIZIO MARIA MAGGI}
\affiliation{%
  \institution{University of Tartu}
  \streetaddress{}
  \city{Tartu}
  \state{}
  \country{ESTONIA}}

\renewcommand{\shortauthors}{Trovato and Tobin, et al.}

%
% The abstract is a short summary of the work to be presented in the article.
\begin{abstract}
It is common for business processes to exhibit a high degree of internal heterogeneity, in the sense that the executions of the process differ widely from each other due to contextual factors, human factors, or deliberate business decisions.
For example, a quote-to-cash process in a multinational company is typically executed differently across different countries or even across different regions in the same country.
Similarly, an insurance claims handling process might be executed differently across different claims handling centres or across multiple teams within the same claims handling centre.
A subset of executions of a business process that can be distinguished from others based on a given predicate (e.g. the executions of a process in a given country) is called a process variant.
Understanding differences between process variants helps analysts and managers to make informed decisions as to how to standardize or otherwise improve a business process, for example by helping them find out what makes it that a given variant exhibits a higher performance than another one.
Process variant analysis is a family of techniques to analyze event logs produced during the execution of a process, in order to identify and explain the differences between two or more process variants.
A wide range of methods for process variant analysis have been proposed in the past decade. 
However, due to the interdisciplinary nature of this field, the proposed methods and the types of differences they can identify vary widely, and there is a lack of a unifying view of the field. 
To close this gap, this article presents a systematic literature review of methods for process variant analysis.
The identified studies are classified according to their inputs, outputs, analysis purpose, underpinning algorithms, and extra-functional characteristics. 
The paper closes with a broad classification of approaches into three categories based on the paradigm they employ to compare multiple process variants.
%The analysis leads to the identification of three broad classes of approaches, which provide a starting point map of the field.
%The article also provides a general methodology for process variant analysis including the required steps to apply this type of analysis for the identification of variants in business process executions.
\end{abstract}

%
% The code below is generated by the tool at http://dl.acm.org/ccs.cfm.
% Please copy and paste the code instead of the example below.
%

% \ccsdesc[300]{Computer systems organization~Redundancy}
% \ccsdesc{Computer systems organization~Robotics}
% \ccsdesc[100]{Networks~Network reliability}

%
% Keywords. The author(s) should pick words that accurately describe the work being
% presented. Separate the keywords with commas.

\maketitle

\section{Introduction}
\label{section:Intro}

Process mining~\cite{vanderAalst:2016:PMD:2948762} is a body of methods and tools to analyze business process execution logs (called \emph{event logs}), in order to extract insights about possible performance deficiencies and improvement opportunities. In this context, an event log is a collection of \emph{traces}, each one consisting of the sequence of events recorded during the execution of one process instance  (herein called a \emph{case}).

Depending on their inputs and their outputs, the following categories of process mining techniques can be distinguished~\cite{DBLP:books/sp/DumasRMR18}:
\begin{itemize}
    \item Automated process discovery techniques, which allow one to discover a business process model from an event log.
    \item Conformance checking techniques, which allow one to compare a process model against an event log in order to qualify and quantify their differences.
    \item Performance mining techniques, which allow one to enhance a given process model with performance information extracted from an event log.
    \item Variant analysis techniques, which allow one to compare two or more event logs corresponding to different variants of a business process, in order to qualify their differences.% and to relate these differences to performance measures.
\end{itemize}

This article deals with the latter category of techniques. The goal of business process variant analysis
%, in general, is to focus on significant and common occurrences of a phenomenon under study to gain insight about exceptional occurrences \cite{Tregear2013}. This idea can be applied effectively to business process management. Indeed, process variant analysis can help 
is to help business analysts to understand \textit{why} and \textit{how} multiple variants of a process differ. In this setting, a \emph{process variant} is 
a subset of executions of a business process that can be distinguished from others based on some characteristic. For example, if a process is executed in three countries, say C1, C2 and C3, we can distinguish three variants of this process: one for each of these countries.
%Similarly, in an insurance claims handling process that deals with multiple types of claims (e.g. home insurance, motor insurance, personal liability insurance, etc.) one can distinguish as many variants as there are types of claims.

Given an event log of a business process, a process variant takes the form of a set of traces (herein called a  \emph{cohort}) that can be separated from others based on a predicate, i.e.\ a function that maps each trace in the log to a boolean variable. The first step in process variant analysis is to split the event log into cohorts using a trace filtering operation. In the above scenario, the predicate that characterizes the first variant is ``country = C1". By applying a log filter that retains only those traces for which this predicate holds, we can extract the cohort corresponding to the first process variant, and similarly for the other two variants.

%One way of creating process cohorts from an event log is to group traces based on a case attribute. In the context of an order-to-cash process, we can for example group traces based on the type of customer, the type of product, or the geographic region of the customer. 

Given that an event log has been split into multiple cohorts, relevant questions that variant analysis seeks to answer include: \textit{why do the executions of a given cohort take longer to complete, on average, than those of another cohort?} Or \textit{what activities are often  skipped in one cohort but are never or seldom skipped in another cohort?} 

As hinted by these questions, variant analysis techniques may cover different perspectives of a business process, including the following ones:
%To answer these questions systematically, process variant analysis can be factorized into the following aspects:
\begin{itemize}
    \item \textbf{Control flow}: Along this perspective, the variants are compared in terms of the occurrence of activities in the execution traces and their relative execution order.
    %It aims at uncovering common patterns in the sequences of activities belonging to different process cohorts.
    \item \textbf{Performance}: Along this perspective, the variants are compared in terms of performance characteristics or performance measures. 
    %The analysis can be conducted for every performance attribute either inside a process cohort, i.e., \textit{inter-cohort}, or among a set of process cohorts, i.e., \textit{intra-cohort}.
\end{itemize}

\begin{figure}[hbtp]
	\centering
	\includegraphics[width=1\linewidth]{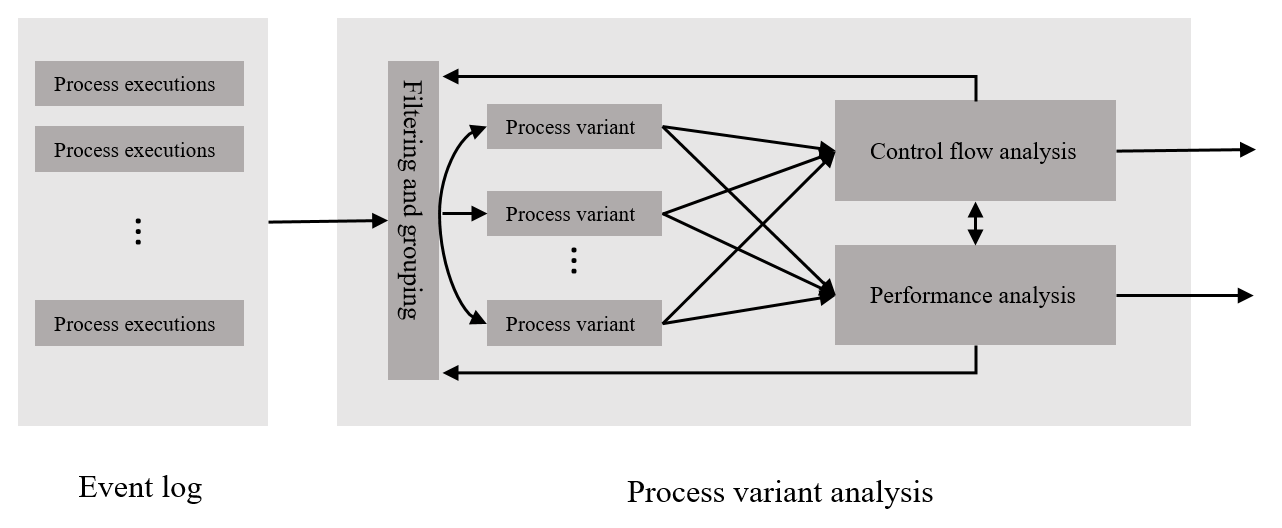}
	\caption{A general framework for process variant analysis}
	\label{fig:general framework}
\end{figure}

The above considerations are depicted in \figurename~\ref{fig:general framework}, which shows that variant analysis starts by splitting an event log into multiple cohorts, which are then compared according to different perspectives, including the control-flow and the performance perspectives.

% determination of attributes Success and Failure measures upon which cases or process executions can be grouped into deviant and non-deviant cases.

% Experiments and recent works in this direction have indicated two kinds of patterns are of interest for researchers. First, a pattern that aims to shed light on the existing differences among the set of process executions, i.e., interpretability. Second, a pattern that aims to distinguish a set of process executions from the others, i.e., discriminatory. Though each kind of pattern is worthwhile, there is a trade-off between interpretability and the discriminatory aspects. The compromise between these aspects shows a negative correlation, i.e., the more interpretable the pattern, the less discriminatory power it has and vice versa.  However, explainable patterns provide a more natural means for the interaction between process analysts and non-technical users.

A wide range of methods for log-based process variant analysis have been proposed in the past decade. 
However, due to the interdisciplinary nature of this field, the proposed methods and the types of differences they can identify vary widely, and there is a lack of a unifying view of the field. 
To close this gap, this article presents a systematic literature review of methods for process variant analysis. The article also proposes a taxonomy of existing methods and identifies gaps in the field.

%Some of the collected techniques have different names in the literature; however, they pursue similar objectives. 

The article is organized as follows. Section \ref{sec:preliminaries} introduces background concepts and terminology used in subsequent sections. Following that, Section \ref{section: search methology} describes the search and selection criteria for identifying relevant studies. Next, Section \ref{sec: analysis and classification} provides an in-depth analysis and detailed classification of the identified studies. Section \ref{sec: methodology framework} presents a broader classification of approaches in terms of the paradigms employed to compare process variants. Finally, Section \ref{sec: conclusion} summarizes the findings.

\section{Preliminaries AND BACKGROUND}
\label{sec:preliminaries}

Process variant analysis, as we will explain in the upcoming sections, has been tackled in two different fields: process mining and machine learning. This section provides basic concepts that will help us to explain how process variant analysis has been approached in each of these the two fields.  

% for those readers with limited exposure to the main concepts. 

\subsection{Process mining}
Process mining is a research area between Business Process Management (BPM) and data science that is concerned with deriving useful insights from process execution data. Process mining techniques can support various phases of the BPM life-cycle, such as process discovery, process analysis and process monitoring \cite{vanderAalst:2016:PMD:2948762}. In fact, it aims at discovering, monitoring and improving real processes by
extracting knowledge from event logs readily available in today's information systems \cite{vanderAalst:2016:PMD:2948762}. The recent significant growth of event data available on the one side and the development of mature process mining techniques on the other side are pushing companies and organizations to exploit process mining to analyze and improve their processes.

The input artifacts for process mining are a \textit{process model} and an \textit{event log}. A process model shows the \textit{expected behaviour} of the process, and the event log shows the process executions, a.k.a. \textit{footprint} or \textit{observed behavior}. Process mining techniques can be classified into three types. The first type, \textit{discovery}, aims at discovering a process model from an event log without using any a-prior information. The second type, \textit{conformance checking}, focuses on confronting an event log and a process model (discovered from an event log or manually designed). Conformance checking is used to check if reality, as recorded in the log, conforms to the model and vice versa. The third type, \textit{Enhancement}, intends to improve an existing process model by using the information about the actual process executions recorded in the event log, or the disconformities identified via conformance checking.

An event log consists of \textit{cases} or \textit{traces}, each capturing a particular execution of a business process. Each case consists of a number of events and each event represents the execution of a particular activity in the process. Each event has a range of attributes of which three are mandatory: i) the case identifier specifying which case generated this event, ii) the event class (or activity name) indicating which activity the event refers to, and iii) the timestamp indicating the completion time of the activity. Note that, in process mining approaches, the completion time of each event determines the order of the events. We call \textit{performance attributes} all the other attributes different from the ones mentioned above.

\begin{table}[h]
\begin{tabular}{llll|llllll}
\hline \hline
\multicolumn{5}{l}{Case attributes} & \multicolumn{5}{l}{Event attributes}             \\
\hline
Id   & City   & Sex   & Product     & Activity & Completion time & Starting time & Resource &   \\
\hline
  1   &   NY     &  M     &   Book     &  Order  &   1/1/2017 9:13:00    &   1/1/2017 9:12:57 &   David         &           &   \\
  1   &   NY    &   M     &   Book     &  Pay in cash  &   1/1/2017 9:14:20  &   1/1/2017 9:14:10     &   John              &           &       \\
  1   &   NY     &  M     &   Book     &  Approval  &     1/1/2017 9:16:00     &     1/1/2017 9:15:37 &  Tiffany            &           &       \\ 
\hline
  2   &   MA    &   F    &    Sofa    &  Order  &       2/1/2017 16:55:00   &  2/1/2017 16:45:00   &        Joe       &           &       \\
  2   &   MA     &  F    &    Sofa    &  Pay by card  &    2/1/2017 17:00:00      &  2/1/2017 16:58:00      &   Nathan            &           &        \\
  2   &   MA     &  F    &    Sofa    &  Disapproval  &   3/1/2017 9:00:00       &   3/1/2017 8:57:00       &    Jane          &           &        \\
  2   &   MA      & F    &    Sofa    &  Pay in cash  &   3/1/2017 9:01:50       &   3/1/2017 9:01:20       &    John          &           &       \\
  2   &   MA     &  F    &    Sofa    &  Approval  &    3/1/2017 9:03:10      &  3/1/2017 9:02:12      &       Luis        &           &      \\
  \hline
  3   &   LA     &  M     &   T.V.     &  Order  &   1/1/2017 9:13:00    &   1/1/2017 9:10:00    &   James         &           &        \\
  3   &   LA    &   M     &   T.V.     &  Pay by card  &   1/1/2017 9:24:20       &  1/1/2017 9:22:20       &   Patrick              &           &        \\
  3   &   LA     &  M     &   T.V.     &  Approval  &     1/1/2017 9:26:00     &      1/1/2017 9:14:00 &    Carla            &           &         \\ 
  \hline
    4   &   LA     &  F     &   Book     &  Order  &   1/1/2017 7:13:00    &   1/1/2017 7:10:00    &   James         &           &        \\
  4   &   LA    &  F     &   Book     &  Pay in cash  &   1/1/2017 7:24:20       &  1/1/2017 7:22:20       &   Patrick              &           &        \\
  4   &   LA     &  F     &   Book     &  Approval  &     1/1/2017 7:26:00     &      1/1/2017 7:25:00 &    Carla            &           &         \\ 
%   \hline
%   \vdots\\
  \hline 
\end{tabular}
\caption{A sample of an event log from an online retailer}
\label{table:event log example}
\end{table}

For example, \tablename~\ref{table:event log example} shows an event log for a simplified online shopping process from a retailer. A case in this table has four (case) attributes, \textit{Id}, \textit{City} (the place where the buyer lives), \textit{Sex} (of the buyer), and \textit{Product}. Also, each event has several (event) attributes such as \textit{Activity}, \textit{Starting time}, \textit{Completion time}, and \textit{Resource} (who processes the activity from the retailer side). The order of activities inside a case is called \textit{control flow}. For instance, in the first case (Id=1), the customer starts by ordering a book (\textit{Order}), then he pays in cash (\textit{Pay in cash}), and finally the retailer approves the payment (\textit{Approval}). 

We now define the mentioned concepts formally.

%\begin{definition}[Event]
%\label{def: event}
% An event is a tuple $(a, c, t , (d_1,v_1), . . . , (d_m,v_m))$, where $a$ is the event name or activity, $c$ is the case identifier, $t$ is the timestamp, and $(d_i, v_i)$ is an event attribute where the first element is the attribute name and the second one is the attribute value.
%\end{definition}

%\begin{definition}[Trace]
%\label{def: trace}
% Let $\Sigma$ be the universe of all activities. A trace is a non-empty sequence $\sigma = \langle a_1, \dots, a_n \rangle$ of events such that
%$\forall i \in  [1,n], a_i \in \Sigma$, $\forall i, j \in [1,n]$,  $a_i .c = a_j .c$, and it does not exist an  (i.e., all events in the trace refer to an entire case).
%\end{definition}

%A trace can be viewed  as a string that represents a finite sequence of activities. A multiset of traces constitutes an event log, formally:

%\begin{definition}[Event log]
%An event log $L$ is a multiset of traces, i.e., $L = {\sigma_i : \sigma_i \in \Sigma^*}$, where $\Sigma^*$ is the set of all possible traces over the universe of activities.
%\footnote{Kleene star operator, i.e., $\Sigma^* = \bigcup_{i \geq 0} \Sigma^0 \cup \Sigma^1 \cup \Sigma^2 \cup \dots $.}
%\end{definition}

\begin{definition}[Event] An $event$ is a tuple $(a, c, t, (d_1, v_1), \ldots, (d_m, v_m))$ where $a$ is the activity name, $c$ is the case id, $t$ is the timestamp and $(d_1, v_1) \ldots, (d_m, v_m)$ (where $m \geq 0$) are the event or case attributes and their values.
\end{definition}
%We call $\mathcal{E} = \mathcal{A} \times \mathcal{C} \times \mathcal{T} \times \mathcal{D}_1 \times \ldots \times \mathcal{D}_m$ the event universe.

\noindent The universe of all events is hereby denoted by $\mathcal{E}$. If we consider \tablename~\ref{table:event log example}, an event is (\textit{Order}, 1, \text{1/1/2017 9:13:00}, (\text{Starting time}, \text{1/1/2017 9:12:57}), (\text{Resource}, \text{David})).

The sequence of events generated by a given process execution forms a \emph{trace}. Formally:
\begin{definition}[Trace] A $trace$ is a non-empty sequence $\sigma = [e_1,\ldots,e_{n}]$ of events such that $\forall i \in [1..n], e_i \in \mathcal{E}$, and $\forall i,j \in [1..n] \; e_i{.}c = e_j{.}c$. In other words, all events in the trace refer to the same case id.
\end{definition}

The corresponding trace for the first case is [(\textit{Order}, 1, \text{1/1/2017 9:13:00}, (\text{Starting time}, \text{1/1/2017} \text{9:12:57}), (\text{Resource}, \text{David})), (\textit{Pay in cash}, 1, \text{1/1/2017  9:14:20}, (\text{Starting time}, \text{1/1/2017  9:14:10}), (\text{Resource}, \text{John})), (\textit{Approval}, 1, \text{1/1/2017  9:16:00}, (\text{Starting time}, \text{1/1/2017  9:15:37}), (\text{Resource}, \text{Tiffany}))].  

A set of \emph{traces} is called an \emph{event log}. Also, we can create process variants based on case attributes such as \textit{Sex, Product}, or the cycle time of a case.

\begin{definition}[Process cohort (or Process variant)]
An event log $L$ can be partitioned into a finite set of groups called process variants or process variants $\varsigma_1, \varsigma_2, \dots, \varsigma_n$, such that $L = \varsigma_1 \cup \varsigma_2 \cup \dots \varsigma_n$, and $\forall i,j, \varsigma_i \cap \varsigma_j = \emptyset$, and, $\exists {d}$ such that $\forall$ $\varsigma_k$ and  $\forall \sigma_i, \sigma_j \in \varsigma_k$, $\sigma_i.d = \sigma_j.d$.
\end{definition}

\noindent The above definition of a process variant emphasizes that the process executions in the same group must share the same attribute value for a given attribute, and each process execution belongs only to one process variant.

%At this point, we stress that through this paper, the following terms, i.e., \textit{process execution, process execution, case} and \textit{trace} are used equivalently unless we explicitly state it. Similarly, \textit{process variants}, \textit{process groups}, and \textit{process cohorts} refer to the same concepts and we will use them interchangeably. Additionally, for \tablename~\ref{table:event log example}, 
%for the sake of simplification, we show \textit{Order} as $a_1$, \textit{Pay in cash} as $a_2$, \textit{Pay by card} as $a_3$, \textit{Approval} as $a_4$, and \textit{Disapproval} as $a_5$.

A process model is a graphical entity used to represent how a process is executed in an organization. In the business domain, a business process model is a collection of inter-related events, activities, and decision points that involve a number of actors and objects, which collectively lead to an outcome that is of value for a customer \cite{DBLP:books/sp/DumasRMR18}.  Companies and organizations usually use different notations to represent their business process models and each of them has different characteristics. Thus, selecting an appropriate process modeling language is essential. However, it is worth mentioning that often one formalism can easily be translated to other notations \cite{vanderAalst2003}. In the following, we present a short introduction to Petri nets \cite{Murata89} and transition systems \cite{vanderAalst:2016:PMD:2948762}, the most used notations to formally represent business process models. 

%\begin{definition}[Petri net]
%A \emph{Petri Net} (PN) \cite{Murata89}, is a 3-tuple $N = \langle P,T,\mathcal{F} \rangle$, where
%$P$ is the set of places, $T$ is the set of transitions,  $P \cap T = \emptyset$, \mbox{$\mathcal{F}: (P \times T)
%	\cup (T \times P) \to \{0,1\}$} is the set of directed arcs called flow relation.
%\end{definition}

% A Petri net with a labeling function $\ell: T\rightarrow \Sigma \cup \{\tau\}$ is called \textit{labeled} Petri net. The labeling function maps transitions to the universal set of activities $\Sigma$. Also, a \textit{marked} Petri net is a pair
%$(N, m)$, where $N = (P , T , F )$ is a Petri net and where $m \in B(P)$ is a multiset over $P$ denoting the marking of the net. A \textit{marking}
%is an assignment of non-negative integers to places. If $k$ is assigned to place $p$
%by marking $m$ (denoted $m[p] = k$), we say that $p$ is marked with $k$ \textit{tokens}.
% The dynamic behavior of such a marked Petri net is defined by the so-called
%\textit{firing rule}. A transition is \textit{enabled} if each of its input places contains a token. An
%enabled transition can \textit{fire} therefore consuming one token from each input place and
%producing one token for each output place. The procedure just mentioned is %called \textit{token replay}.

A \emph{Petri net} $N=(P,T,F)$ is a directed graph with a set $P$ of nodes called \emph{places} and a set %second
$T$ of \emph{transitions}. Places are represented by circles and transitions by squares. The nodes are connected via directed arcs $F\subseteq (P \times T) \cup (T \times P)$. Connections between two nodes of the same type are not allowed. 
Given a transition $t \in T$, ${}^{\bullet} t$ is used to indicate the set of \emph{input places} of $t$, which are
the places $p$ with a directed arc from $p$ to $t$ (i.e., such that $(p,t) \in F$).
Similarly, $t^\bullet$ indicates the set of \emph{output places}, namely the places $p$
with a direct arc from $t$ to $p$.
At any time, a place can contain zero or more \emph{tokens}, drawn as black dots. The state of a Petri net, a.k.a. \emph{marking} $m$, is determined by the number of tokens in places, i.e., $m: P \rightarrow \mathsf{N}$.

In any run of a Petri net, the number of tokens in places (i.e., the marking) may change.
A transition $t$ is \emph{enabled} at a marking $m$ iff each input place contains at least one token, i.e., $\forall$ $p \in {}^{\bullet} t$, $M(p)>0$. A transition $t$ can \emph{fire} at a marking $m$ iff it is enabled. As result of firing a transition $t$, one token is ``consumed'' from each input place and one is ``produced'' in each output place.
This is denoted as $m \xrightarrow{t} m'$. 
%In the remainder, given a sequence of transition firing $\sigma=\langle t_1,\ldots,t_n\rangle \in T^*$, $m_0\xrightarrow{\sigma} m_n $ is used to %compactly indicate $m_0 \xrightarrow{t_1} m_1 \xrightarrow{t_2} \ldots \xrightarrow{t_n} m_n$.

\begin{figure}[h]
	\centering
	\includegraphics[width=.5\linewidth]{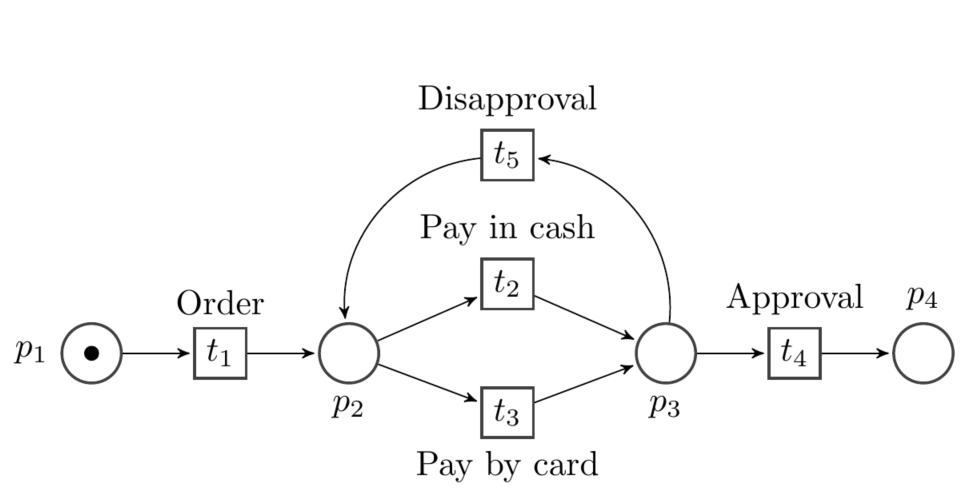}
	\caption{Labeled Petri net, transitions are squares, places are circles and tokens are black dots }
	\label{fig:labled petri net}
\end{figure}

For example, consider the process model in \figurename~\ref{fig:labled petri net} reflecting the behavior of the event log in \tablename~\ref{table:event log example}. The set of transitions and places are $\{t_1, t_2, t_3, t_4, t_5\}$ and $\{p_1, p_2, p_3, p_4\}$, respectively. Also, the labeling function is $\ell(t_1)= \text{``Order''},~ \ell(t_2)= \text{``Pay in cash''},~ \ell(t_3)= \text{``Pay by card''},~ \ell(t_4)= \text{``Approval''},~ \ell(t_5)= \text{``Disapproval''}$. In the process model, only $p_1$ has one token, i.e., $m[p_1]=1$, moreover, $t_1$ is enabled and ready to fire. To show how the model executes, suppose that $t_1$ fires, then it consumes one token from $p_1$ and produces one token into $p_2$, thus, $t_2$ and $t_3$ become enabled; however, only one of them can fire. After firing $t_2$ or $t_3$, then one token is placed in $p_3$, which enables $t_4$ and $t_5$. Finally, one of $t_5$ or $t_4$ is fired, where the former marks $p_2$ and the execution continues, whereas the latter marks $p_4$ and the execution terminates.

\begin{figure}[h]
	\centering
	\includegraphics[width=.5\linewidth]{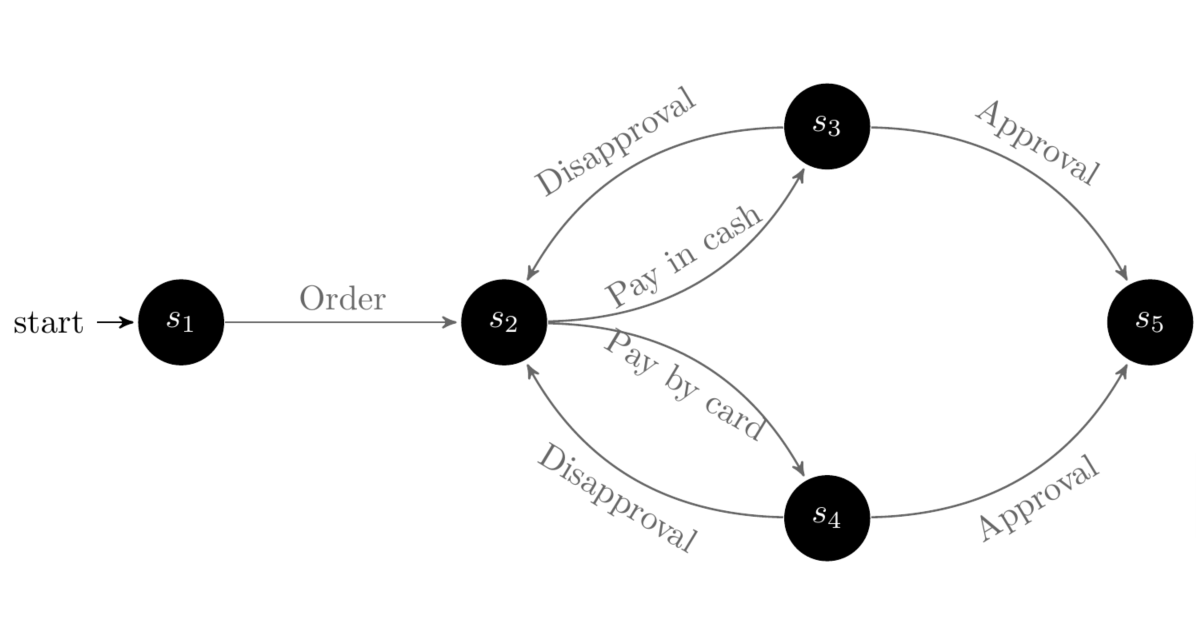}
	\caption{A transition system having one initial state and one final state }
	\label{fig:transition system}
\end{figure}

A \textit{transition system} is a triplet $TS = (S,A,T)$, where $S$ is the set of \textit{states}, $A \subseteq \Sigma$ is the set of \textit{activities} (often referred to actions), and $T \subseteq S \times A \times S$ is the set of \textit{transitions}. $S^{start} \subseteq S$ is the set of \textit{initial
states}, and $S^{end} \subseteq S$ is the set of \textit{final states}.
A transition system is the most basic process modeling formalism compared to other notations; it is also known as a \textit{Directed Graph (DG)}. As an example, consider the transition system in \figurename~\ref{fig:transition system} reflecting the behavior of the event log in \tablename~\ref{table:event log example}. The corresponding set of states and activities are $S=\{s_1, s_2,s_3,s_4,s_5\}$, and $A=\{\textit{Order, Pay in cash, Pay by card, Disapproval}$, $\textit{Approval}\}$. Also, $S^{start} = s_1$, and $S^{end} = s_5$.

Two important concepts that would be helpful in variant analysis of process executions are the notions of \textit{replaying} \cite{vanderAalst:2016:PMD:2948762} and \textit{alignment} \cite{Adriansyah.Thesis.2014}. Replaying a process execution on a process model means to rerun the process execution on the process model to quantify discrepancies between them. Though replaying provides useful and easy-to-understand information, a more fundamental way to identify such deviations is by using alignments. 
Alignments play an important role in conformance checking. Given a process model and a process execution, an alignment quantifies to what extent the process model can mimic the process execution. An alignment is a two-row matrix that lines up corresponding activities in the process model and in the process execution. Formally:

\begin{definition}[Alignment]
\label{def:alignment}
Given a process model and a process execution, let $\Sigma$ be the universe of all activities. Let $A_M \subseteq \Sigma$ and $A_L \subseteq \Sigma$ be the alphabet of activities in the model and events in the event log, respectively, and $\perp$ the empty set, then an alignment, denoted by $\alpha$, is a sequence of \textit{legal} moves, where:
	
	\begin{itemize}
		\item $(x,y)$ is a \textit{synchronous move} if $x \in A_L$, $y \in A_M$
		\item $(x,y)$ is a \textit{move in log} if $x \in A_L$ and $y= \perp$.
		\item $(x,y)$ is a \textit{move in model} if $x= \perp$ and $y \in A_M$.
		\item $(x,y)$ is an \textit{illegal move},  otherwise.
	\end{itemize}
% 	The set of all \textit{legal} moves is denoted as $A_{LM}$ and given an alignment $\alpha \in A_{LM}^*$,
% 	the projection of the first element 
% 	(ignoring $\perp$), $\alpha \downarrow _{A_L}$, results in the observed 
% 	trace $\sigma$, and 
% 	projecting the second element (ignoring $\perp$), $\alpha \downarrow _{A_M}$, results in the model trace.
\end{definition}

For example, an alignment between the process execution $\sigma_L =$ [Order, Approval, Pay by card], and the process model in \figurename~\ref{fig:labled petri net}, with initial marking and final marking denoted with $m_i$ (a single token in $p_1$) and $m_f$ (a single token in $p_4$), is the following:

\begin{center}
  $\alpha$=\begin{tabular}{ | c | c | c | c | c | r |}
	%\hline			
	Order & Approval  & Pay by card  &$\perp$   \\ 
	\hline
	Order & $\perp$ & Pay by card & Approval      \\ 
\end{tabular}  
\end{center}

\noindent In this example (Order, Order) and (Pay by card, Pay by card) are synchronous moves, and (Approval, $\perp$) and ($\perp$, Approval) are move in log and model respectively, or, in short, \textit{asynchronous} moves. 
Note that, ignoring all occurrences of $\perp$, the projection on the first element of the moves yields $\sigma_L$ and the projection on the second one yields a sequence $\sigma''$ such that $m_i \xrightarrow{\sigma''} m_f$.
Generally speaking,  a move in log for a transition $t$ indicates that $t$ occurred when not allowed; a move in model for a transition $t$ indicates that $t$ did not occur, when, conversely, expected. 
An alignment usually is quantified with a \textit{fitness} value, which, in the simplest case, is the number of synchronous moves divided by the total number of moves. For the mentioned example, the fitness is $\frac{2}{4}$.

%As this example shows, the model can not imitate the order of activities entirely, i.e., control flow, which happened in the process execution. There could be several reasons that give rise to this misbehavior; however, having asynchronous moves is the sign of deviation that opens a door for further investigations.

\subsection{Machine learning}
\label{subsec: machine learning}
Machine learning is the systematic design, analysis and study of algorithms and systems that learn from past experiences. Machine learning is inherently a  multidisciplinary field. It  draws on results from \textit{artificial intelligence, probability} and \textit{statistics, computational complexity theory, control theory, information theory, philosophy,  psychology,  neurobiology},  and other fields \cite{Mitchell:1997:ML:541177}. 

Given a problem at hand, the first step in learning from data is to have related \textit{observations}. The raw observations comprise \textit{multidimensional data, event log data, graph data}, and other types of data. Moreover, for every type of data, several sophisticated machine learning algorithms have been proposed by researchers. 
However, because of historical and technical reasons, most of the developed algorithms use multidimensional data or encode other types of data into a multidimensional representation. In an n-dimensional representation, every entity is shown as a \textit{vector} of length $n$, and each dimension is called a \textit{feature} or \textit{attribute}. Thus, a group of observations $D$ can be shown as a multiset of vectors as follows:
\begin{equation}
    D = \{\mathbf{x_i}\}_1^m \quad \text{or} \quad \{\mathbf{x_1},\mathbf{x_1}, \dots, \mathbf{x_m}\},~\text{where}~ \forall i \in [1..m], ~\mathbf{x_i} = (x_{i,1}, x_{i,2}, \dots, x_{i,n})
\end{equation}

\noindent In the above representation, $\mathbf{x_i}$ is a vector with $n$ features $x_{i,1}, x_{i,2},\dots, x_{i,n}$. A feature can be a complex structured object, such as an image, a
sentence, a time series, a molecular shape, a graph, a sequence prefix \cite{Murphy:2012:MLP:2380985}.

Broadly speaking, a machine learning task can be of two types:
\begin{itemize}
    \item In \textit{descriptive} or \textit{unsupervised learning} approaches, given a set of observations $D$, the objective is to find interesting patterns in the data. A canonical example of unsupervised learning is the problem of \textit{clustering} data observations into groups.
    
    \item In \textit{supervised learning} or \textit{predictive} approaches, each vector $\mathbf{x_i}$ has an associate label $y_i$, which is called \textit{response variable}. Response variables can be of different nature, but the most methods assumes that it is \textit{categorical} or \textit{real-valued}. The set of labeled vectors, i.e., $D=\{(\mathbf{x_i},y_i)\}_1^m$ is called the \textit{training set}, and the main objective of supervised learning algorithms is to estimate a mapping function from $\mathbf{x}$ to $y$, i.e., $y = f(\mathbf{x})$. The estimated function or the \textit{trained model} is called a \textit{classification} model for categorical response variables, and a \textit{regression} model for real-valued response variables.
\end{itemize}

There exist many well-developed and dedicated algorithms for the machine learning approaches just mentioned. For example, \textit{decision tree} and \textit{rule-based} algorithms and their variants are among the first proposed supervised learning algorithms. A decision tree, using a set of hierarchical decisions on the features, constructs a tree-like structure to classify an input observation. Similarly, a rule-based classifier uses a set of ``if-then'' rules to match \textit{antecedents} to \textit{consequents}. A rule is expressed as follow:
\begin{align}
        \text{IF} \quad Antecedent \quad \text{THEN} \quad Consequent
\end{align}

\noindent where the antecedent is a logical combination of features, e.g., $(x_{i,1} \land x_{i,2}) \lor x_{i,4}$, and the consequent is the class label. Rule-based algorithms are the supervised version of \textit{association rule mining} algorithms, which determine relationships in a set of observations.

Though decision tree and rule-based classifiers adopt different underlying mechanisms for classification tasks, a decision tree may be viewed as a particular case of a rule-based classifier in which each
path of the decision tree corresponds to a rule. 
%Having said that, one can summarize a decision tree via a set of rules.

From the probabilistic perspective, despite the variety of proposed supervised and unsupervised learning algorithms, either try to approximate probability values. In particular, supervised learning algorithms strive to approximate $p(y_i | \mathbf{x_i})$, i.e., the probability of a class label given an input vector, whereas an unsupervised algorithm can be viewed as a \textit{density estimation}, i.e., $p(\mathbf{x_i})$ \cite{Murphy:2012:MLP:2380985}. The differences among machine learning algorithms are in the way they compute these probabilities.

The performance of a machine learning algorithm can be evaluated in different ways. For unsupervised learning algorithms, the validation is often difficult since the problem is defined in a descriptive way. However, some validation criteria can be defined to evaluate the objective function upon which observations are clustered together. In contrast, the predictive ability of a supervised learning algorithm can be evaluated using the input labels.
For example, \textit{accuracy} and \textit{Area Under Curve} (AUC) can be used to evaluate a classification model. The former shows the ratio of the number of correct predictions to the total number of predictions, and the latter, for a binary classification model, provides the probability that a model ranks a random positive example more highly than a random negative example.

A learning paradigm that has received much attention over the past few years is the \textit{learning by committee} or \textit{ensemble learning} \cite{Dietterich:2000:EMM:648054.743935}. Ensemble learning is motivated by the fact that, given a problem, different learning algorithms might provide different results due to the specific characteristics of the underlying learning algorithms, or their sensitivity to the random artifacts in the input. Therefore, the goal of ensemble learning is to combine the results from multiple learners to improve the quality of the results. In unsupervised learning approaches, it is evident that there are many alternative solutions, i.e., clustering models, alongside a large number of validation criteria, and no single model or validation criterion provides the optimal clustering. Thus, \textit{ensemble clustering}, proposed by \cite{Strehl:2003:ClusterEnsemble:944919.944935}, combines many clustering models to create a more robust clustering approach. By the same token, in supervised learning, a set of \textit{base learners} is created and trained in different ways, and then the results of base learners are combined to create the final prediction. A very simple way to combine outputs of base learners, for real-valued outputs, is to average them:
\begin{align}
    f(x) = \frac{1}{p} \sum\limits_{j=1}^p  f_j (x) 
\end{align}

\noindent In the above expression, there are $p$ base learners, and  $f_j (x)$ is the output of the $j$-th base learner. 

\begin{figure}[h]
	\centering
	\includegraphics[width=.8\linewidth]{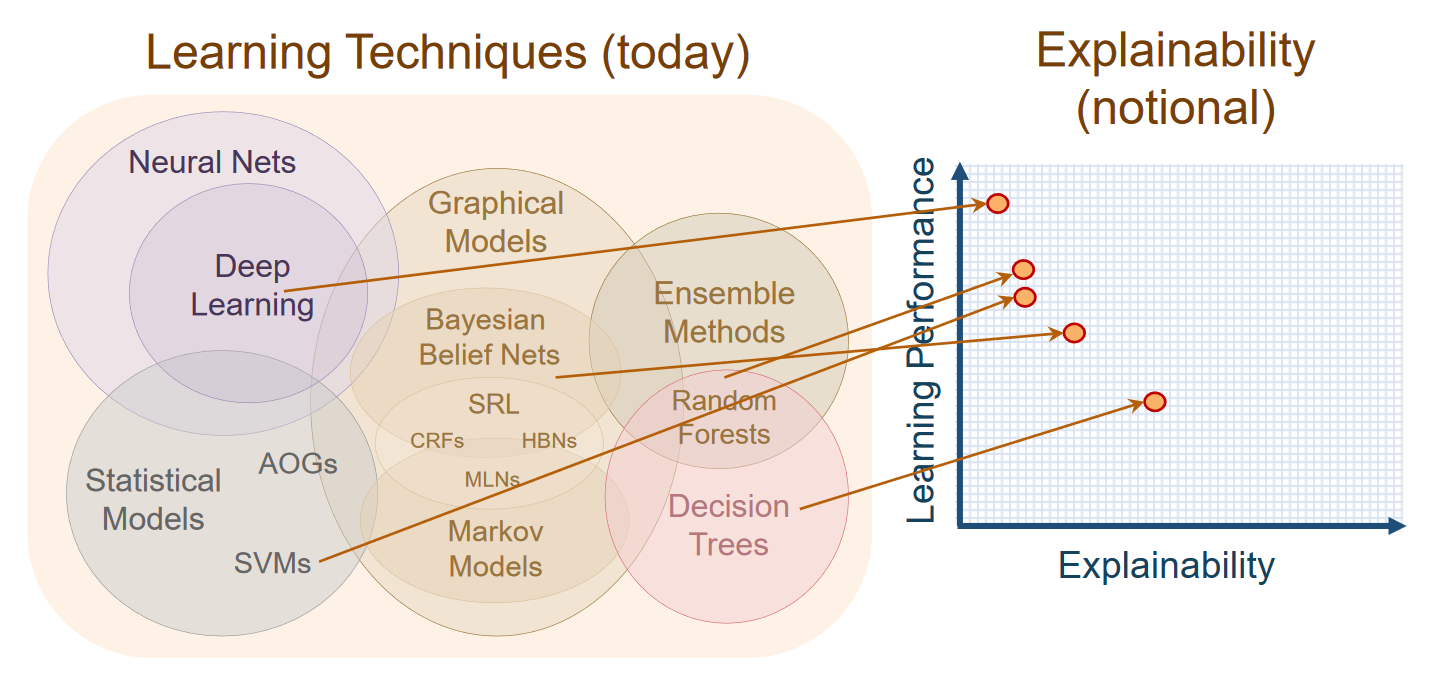}
	\caption[Caption for LOF]{Trade-off between accuracy and explainability of learning algorithms\footnotemark}
	\label{fig:accuracy-explainability}
\end{figure}
\footnotetext{From: \url{https://www.darpa.mil/attachments/XAIProgramUpdate.pdf?source=post_page}}

Notwithstanding the importance of the accuracy of a machine learning algorithm, an algorithm can also be evaluated from other perspectives. For example, in several situations, it is necessary to have an \textit{explainable} machine learning model. Explainability is defined as the science of comprehending what a model did, or might have done \cite{Leilani2018}. More simply, explainability is the extent to which the internal mechanics of a machine learning system can be explained in human terms. The concept of explainability can be applied to all supervised and unsupervised learning approaches. For example, decision tree and rule-based classifiers are highly explainable, i.e., the internal structure of a decision tree and a set of rules can be easily explained in human terms; on the other hand, the internal structure of ensemble models is very difficult to grasp in human terms. Although both accuracy and explainability are two important aspects of a machine learning algorithm, they interfere with each other. Indeed, the internal structure of a sophisticated machine learning algorithm that comes up with very high accuracy is hardly explainable in human terms and it acts as a \textit{black-box}. In this sense, according to the \textit{no free lunch theorem}, there is no universal best model \cite{Wolpert:1997:NFL:2221336.2221408}.
Figure \ref{fig:accuracy-explainability} presents the trade-off between accuracy and explainability aspects for well-known machine learning algorithms.

\section{Search Methodology}
\label{section: search methology}
We conducted a \textit{Systematic Literature  Review} (SLR) of process variant analysis methods, by following the SLR guidelines in~\cite{Kitchenham07guidelinesfor}. In line with these guidelines, we started by posing a research question to clarify the goals of the search. From the research question, a search string was derived for retrieving related documents from academic digital libraries. The following subsections detail the SLR steps followed in this paper.

\subsection{Research Question}
\label{subseq: research question}
The main aim of this paper is to review proposed methods for process variant analysis. Process variant analysis is a rather broad topic. Therefore, to confine our search space, we defined the following research question (RQ): \emph{Given a set of event logs of two or more variants of process, how to identify and explain the differences among these variants?}
% \noindent To approach this question, we decompose it into two sub-questions: 
% \begin{itemize}
%     \item $RQ_1$: How to identify the existing differences between two event logs?
%     \item $RQ_2$: How to explain the existing differences between two event logs?
% \end{itemize}

\subsection{Study Retrieval and Selection}
To retrieve relevant papers based on RQ, the following keywords were considered:
\begin{itemize}
    % \item "(business) process" - a relevant study must be related to the executions of (business) process models;
    % \item "deviance mining" - a relevant study should concern to identify the deviations between event logs of the same process model;
    % \item  "log analysis" - a relevant study should concern to analyze and summarize attributes of an event log;
    % \item  "sequence mining" - a relevant study should concern to identify patterns for an event log;
    \item ``event log'' - a relevant study must consider event logs as inputs;
    \item ``process variant analysis'' - a relevant study should concern the analysis of the executions of a process;
    \item ``process variants comparison'' - a relevant study should concern the comparison of sets of process executions;
\end{itemize}

%  The last keyword, i.e., "sequence mining", was selected to broaden the search result as to embody works from other disciplines like machine learning and bioinformatics as they tackle a rather similar problem with different approaches.
%  Eventually, the following search phrases are considered, 
% "process deviance mining", "business process sequence mining", "process log analysis" and "sequence deviance mining".
Though the aforementioned terms are the most related keywords, we realized that some works related to process variant analysis use the term ``deviance mining'' to indicate this type of analysis; therefore, we included additional terms, namely, ``process deviance mining'' and ``process deviance comparison'', to cover such works.  

Using these keywords we derived a search string that was submitted to Google Scholar. Google Scholar is the world's largest academic search engine, which encompasses other academic databases like ACM Digital Library and IEEE Xplore \cite{Gusenbauer2019}. The retrieved documents are those that have at least one of the above terms in their title, keywords or the main body of the paper. 

\begin{figure}[h]
    \centering
    \includegraphics[width=.7\linewidth]{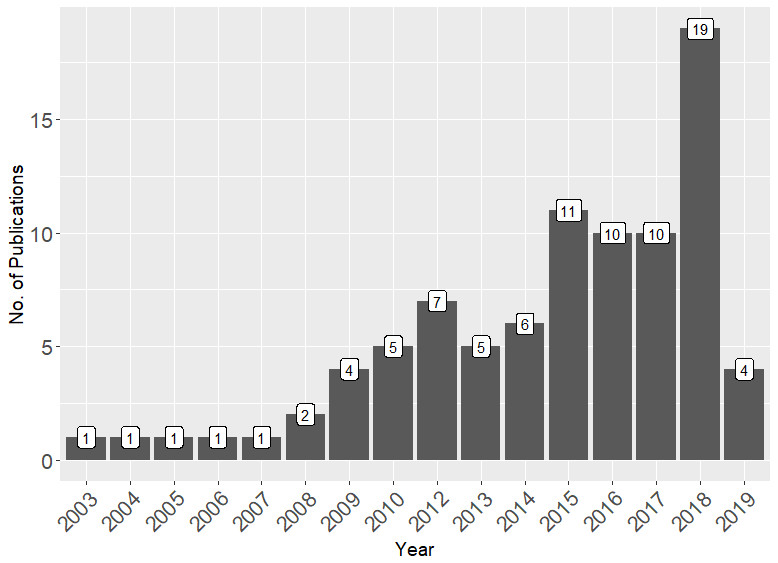}
    \caption{Number of publications per year}
    \label{fig:No-of-publications}
\end{figure}

The search resulted in 88 unique articles published between January 2000 and April 2019. Figure \ref{fig:No-of-publications} shows the number of publications per year according to the proposed search query. One can see an upward trend for the research publications on process variant analysis. This shows that this area of research is recently getting more and more attention.

To eliminate irrelevant results and to avoid exploring marginal studies without any follow-up, we applied the following inclusion criteria:
\begin{itemize}
    \item $INC_1$: The study is about variant analysis of processes (this criterion was assessed by reading title and abstract).
    \item $INC_2$: The study is cited at least five times (this threshold was relaxed for publications from 2018 onward where instead of considering the number of citations we considered, as criterion, the number of pages, i.e., to have at least ten pages single-column or five pages double-column).
\end{itemize}

\noindent We intentionally kept $INC_1$ open by using only the term ``process''. In this way, we can cover different types of processes such as business processes and software development processes.

After applying the above inclusion criteria, we obtained 14 relevant studies. To increase the sensitivity of our research, we proceeded with the Snowball sampling method \cite{SnowballSampling}, i.e., we retrieved the papers that are related to (cite or are cited by) these 14 studies and re-applied the same inclusion criteria as above.
% At this point, it must be mentioned that to cover very recent but relevant works which couldn't pass $INC_2$, a complimentary inclusion criterion is considered as follow:
% \begin{itemize}
%     \item $INC_2^c$: The study is at least ten pages single-column or five pages double-column.
% \end{itemize}
\noindent This procedure resulted in 363 papers, of which we retained 91 unique papers after re-applying the inclusion criteria.

The list of studies that passed the inclusion criteria were further assessed according to a number of exclusion criteria: %Determining if the exclusion criteria are satisfied could require a deeper analysis of the studies, e.g., examining the approach and/or results presented in the paper. The applied exclusion criteria are:
\begin{itemize}
    \item $EX_1$ The study does not propose a concrete technique for comparison of process variants. 
    \item $EX_2$ The proposed technique focuses on building predictive models that can generate predictions based on running process instances, as opposed to supporting the (post-mortem) comparison of process variants.
    \item $EX_3$ The technique does not take an event log as input.
\end{itemize}

More precisely, $EX_1$ excludes those works that are not related to proposing a method for analyzing or comparing process variants. The second exclusion criteria $EX_2$ eliminates works that are focused on predictive process monitoring techniques. The main focus of these latter studies is on predicting future states of ongoing cases, rather than comparing characteristics of sets of completed cases. In addition, predictive monitoring techniques have been studied extensively in previous surveys~\cite{TeinemaaDRM17, Verenich2018SurveyAC, Chamorro2018, DBLP:conf/bpm/Francescomarino18}. The last exclusion criteria $EX_3$ leaves out those studies that do not use event logs as input. These might be studies that compare process models represented using different formalism. %However, those works that compare process execution models, i.e., discovered from event log, of the same process model are considered.
Though these approaches might be inspiring for process variant analysis, the scope of this paper is limited to review the current existing techniques that leverage process executions.
The application of the exclusion criteria resulted in 29 relevant studies out of 91 works selected in the previous step.\footnote{All the retrieved papers can be found at \url{ https://figshare.com/articles/Selected_articles/9999887}}

\section{ANALYSIS AND CLASSIFICATION OF METHODS}
\label{sec: analysis and classification}
Research question RQ can be answered by categorizing the selected works using different dimensions specifying the typology of the existing methods and their characteristics. In particular, each study can be decomposed into the following dimensions:

\begin{itemize}
    \item Input data
    \item Outcome
    \item Process perspective (control flow, resources, data)
    \item Family of algorithms (the main algorithm used in the study)
    \item Evaluation data (real-life or artificial logs) and application domain (e.g., insurance, banking, healthcare)
    \item Implementation (standalone or plug-in, and tool accessibility)
\end{itemize}

\tablename~\ref{table: summary dimension of relevant studies} provides an overview of the identified studies  according to the mentioned dimensions. In the following, we provide an overview of each study and, then, more details about the classification for each dimension. 
%Note that the classification of each work is based on the methods and experiments provided in the analyzed papers, but this does not mean that the work cannot be extended to other dimensions a.

\subsection{Overview}
\label{subsec: overview}

According to our results, the work in \cite{Poelmans2010} is the first work that considers process variant analysis at the process execution level. A process execution, in this work, contains treatment activities that a hospital applies to  breast cancer patients. This work aims at gaining a deeper understanding of an existing breast cancer care process to discover process inefficiencies, exceptions and variations, and to find their root causes. To this end, Hidden Markov Models are used for process discovery and Formal Concept Analysis \cite{Ganter:1997:FCA:550737} is employed to analyze clusters of patients identified in the discovered processes. 

Similarly to this work, a series of interactive tools for extracting and visualizing clinical care pathways is presented in \cite{Lakshmanan:2013:ICC:2529737.2529772}. The work considers a process execution as a sequence of clinical activities that patients receive in their care journeys. The main objective of the paper is to examine the impact and correlation of clinical activities on the clinical care pathway of a patient for specific diseases. Different techniques like frequent pattern mining and trace clustering are applied to accomplish this goal.  In this study, a tool for visualizing the results of the analysis is also presented. The tool discovers dependency graph models using the Heuristic Miner \cite{Weijters2006}, and then the impactful patterns obtained from frequent pattern mining are superimposed to them to highlight differences among different variants.

%\begin{sidewaystable}

\begin{landscape}
\begin{table}[p]
 \scriptsize{

%  \begin{tabular}{p{2cm} | p{.6cm} | p{2.5cm} | p{3.2cm} | p{3.3cm} | p{1.2cm} | p{1.9cm}| p{3cm}}
  \begin{tabular}{p{2.3cm} | p{.6cm} | p{2.5cm} | p{3.7cm} | p{4.1cm} | p{1.5cm} | p{1.4cm}| p{3cm}}
\hline \hline
 Study & Year & Input data & Outcome & Algorithm & Domain & Implement. & Type of analysis \\  
 \hline\hline
 Poelmans et al. \cite{Poelmans2010} & 2010 & Event log & Rule, Descriptive statistics  & Hidden Markov Model, Formal Concept Analysis & Healthcare & Matlab & Control flow, Cycle time\\
 
 \hline
 Swinnen et al. \cite{Swinnen2012} & 2012 & Event log, Process model & Rule &  Fuzzy mining, Association rule mining & Financial & Weka, ProM & Control flow\\
 
 \hline
Buijs et al. \cite{Buijs2011} & 2012 &  Event log  & Alignment matrix, Descriptive statistics &  Alignment analysis &  Public administration & ProM & Control flow\\

 \hline
Suriadi et al. \cite{Suriadi2013} & 2013 & Event log & Rule, Process model (TS), Descriptive statistics &  Fuzzy mining, Causal relation analysis & Insurance & Weka, ProM, Disco & Control flow\\

 \hline
Sun et al. \cite{Sun:2013:MER:2486046.2486067} & 2013 & Event log & Rule &  Contrast itemset mining & Financial, Industrial & MOSPER & Control flow\\

\hline
Bose et al. \cite{Bose2013} & 2013 & Event log & Rule &  Decision tree induction, Association rule mining &  Industrial & ProM & Control flow\\

\hline
% Lakshmanan et al. \cite{Lakshmanan:2013:ICC:2529737.2529772} & 2013 & Event log & Frequent pattern, Process model (TS) &  Frequent pattern mining, DBSCAN &  Helathcare & BPI & Workflow\\

Lakshmanan et al. \cite{Lakshmanan:2013:ICC:2529737.2529772} & 2013 & Event log & Rule, Process model (TS) &  Frequent pattern mining, Clustering &  Healthcare & BPI & Control flow\\

\hline
Kriglstein et al. \cite{Kriglstein2013} & 2013 & Event log,  Process model & Annotated process model (DG) &  Difference Model analysis&  Logistic & N/A & Control flow\\

\hline
% Suriadi et al. \cite{Suriadi2014} & 2014 & Event log &  Variants control flow (PN), Descriptive statistics &  Fuzzy miner, K-means &  Healthcare & RapidMiner, Weka, ProM & Workflow, Waiting time between activities\\

Suriadi et al. \cite{Suriadi2014} & 2014 & Event log &  Process model (PN), Alignment matrix, Descriptive statistics &  Fuzzy mining, Clustering &  Healthcare & RapidMiner, Weka, ProM & Control flow, Waiting time between activities\\

\hline
Buijs et al. \cite{Buijs2014} & 2014 & Event log, Process model  & Alignment matrix, Descriptive statistics & Alignment analysis &  Public administration & ProM & Control flow\\

\hline
Partington et al. \cite{Partington:2015:PMC:2677016.2629446} & 2015 & Event log & Annotated process model (BPMN, TS) & Fuzzy mining, Log replay &  Healthcare & ProM, Nitro & Control flow, Waiting time between activities, Cycle time\\

\hline
Cordes et al. \cite{Cordes2015} & 2015 & Event log & Annotated process model (DG)  & Difference Model analysis (using TGraph) &  Healthcare & Standalone & Control flow\\

\hline
Pini et al. \cite{Pini2015} & 2015 & Event log, Process model & Annotated process model (TS), Alignment matrix, Descriptive statistics & Alignment analysis &  Healthcare & ProM & Control flow, Activity duration, Cycle time \\

\hline
Bolt et al. \cite{Bolt2015} & 2015 & Event log & Annotated process model (TS), Descriptive statistics & Transition system mining &  Education & RapidProM, RapidMiner & Control flow  \\

\hline
Conforti et al. \cite{Conforti2015} & 2015 & Event log & Annotated process model (C-BPMN) & C-BPMN mining, Log replay &  Financial & Apromore & Control flow, Cycle time  \\

\hline
Beest et al. \cite{Beest2015} & 2015 & Event log & Rule (as natural language statement) & Prime Event Structure, Partial Synchronized Product &  Synthetic, Healthcare & Apromore & Control flow   \\

\hline
Cuzzocrea et al. \cite{CuzzocreaFGP16} & 2016 & Event log & Ensemble classifier & Ensemble learning via stacking &  Healthcare & Weka, Standalone, ProM & Control flow, Cycle time   \\

\hline
Bolt et al. \cite{Bolt2016} & 2016 & Event log & Annotated process model (TS) & Transition system mining &  Public administration & ProM & Control flow, Elapsed time   \\

\hline
Andrews et al. \cite{Andrews2016} & 2016 & Event log & Annotated process model (C-BPMN) &  C-BPMN mining, Log replay &  Healthcare & Apromore & Control flow, Cycle time   \\

\hline
Cuzzocrea et al. \cite{Cuzzocrea2017DevianceDiscovery} & 2017 & Event log & Annotated process model (TS) & Transition system mining, Clustering &  Logistic & N/A & Control flow, Cycle time   \\

\hline
Cuzzocrea et al. \cite{CuzzocreaFGP17a} & 2017 & Event log & Ensemble classifier & Ensemble learning via stacking &  Healthcare & N/A & Control flow, Cycle time   \\

\hline
Cuzzocrea et al. \cite{Cuzzocrea2017ExtensionsAA} & 2017 & Event log & Ensemble classifier & Ensemble learning via stacking &  Healthcare & N/A & Control flow, Cycle time   \\

\hline
Folino et al. \cite{Folino:2017:DCA:3019612.3019660} & 2017 & Event log & Rule, Process model (TS) & Clustering, Rule mining, Fuzzy mining &  Logistic & N/A & Control flow, Cycle time   \\

\hline
Wynn et al. \cite{WYNN201793} & 2017 & Event log, Process model &  Annotated process model (PN), Alignment matrix, Descriptive statistics & Alignment analysis, Log replay  &Insurance & ProM & Control flow, Waiting time between activities   \\

\hline
Gulden et al. \cite{Gulden2017} & 2017 & Event log &  Rhythm-eye view & Mining and configuring rhythm-eye visualization  &Industrial & Standalone & Control flow, Waiting time between activities   \\

\hline
Ballambettu et al. \cite{Ballambettu2017} & 2017 & Event log &  Annotated process model (TS) & Process map mining  &Logistic & ProM & Control flow, Waiting time between activities   \\

\hline
Folino et al. \cite{FolinoEnsemble2018} & 2018 & Event log & Ensemble classifier & Ensemble learning via stacking &  Healthcare & Weka & Control flow, Cycle time   \\

\hline
Bolt et al. \cite{bolt2018} & 2018 & Event log & Annotated process model (TS), Rule & Transition system mining, Decision tree induction &  Public administration & ProM & Control flow, Elapsed time   \\

\hline
Nguyen et al. \cite{Nguyen2018-Multiperspectquteprints117962} & 2018 & Event log & Matrix-based representation of Differential Graph & Perspective and Differential Graph & Public administration,  IcM & ProM, Apromore & Multiple perspectives  \\
 \hline\hline
\end{tabular}}
\caption{Overview and summary of the selected works}
\label{table: summary dimension of relevant studies}
\end{table}
\end{landscape}

%\end{sidewaystable}

Another study that examines patient flow variations is presented by Suriadi et al.  \cite{Suriadi2014}. Patient flows include sequences of activities executed both in the Emergency Department (ED) and in the ward.
The study aims at explaining event log variations across four different hospitals.
% In other words, event logs are explored and analyzed in different ways from both process mining and data analysis stances to capture existing variations.
To this end, the comparison of patient flows is done by discovering process models using the Fuzzy Miner \cite{Gunther2005}, and the Heuristic Miner \cite{Weijters2006} using the four hospital sublogs. Then, a Petri net is derived from each discovered model, and its fitness is measured by aligning it with the process executions of the other sublogs (i.e., cross-validation) using the technique presented in \cite{Adriansyah.Thesis.2014}. 
Also, the authors conducted some descriptive analysis such as computing the maximum time for a patient to be discharged from ED across different hospitals to provide more insights about patient flow variations and the corresponding performance. Another work by Suriadi et al. \cite{Suriadi2013} aims at improving the customer satisfaction of a company by reducing the processing time of its business processes. In particular, it tries to improve lengthy process executions, which, instead, are supposed to be fast and simple.
The proposed approach employs a technique called Delta-Analysis. The same technique has been applied also in \cite{Partington:2015:PMC:2677016.2629446} to carry on Root Cause Analysis (RCA) for some specific process executions that take an unexpectedly long time to complete. RCA examines the existing causal relations between various factors that contribute to the execution time of a case via classification algorithms.

Pini et al. \cite{Pini2015} apply some visualization techniques to tackle process variant analysis. The work provides a comparative process visualization technique to compare both performance and control flow of different process variants. The comparison is done using three perspectives, i.e., general model, superimposed model, and side-by-side comparison. Factors such as frequency of an activity and min/max/avg activity durations are used as objective measures to uncover differences among process variants. The general model perspective aims at emphasizing the performance differences among various process variants. The super imposed model perspective draws attention to process flows (i.e., activity ordering) by computing alignments \cite{Adriansyah.Thesis.2014}. The last perspective shows the waiting time between an activity and its successor, thus uncovering which activities inject delays in the whole process execution time. The work in \cite{WYNN201793} proposes an extension of the previous work by considering a normative process model alongside with event logs as inputs, and adding more data preparation facilities. It also provides comparative process visualizations at different levels-of-detail to improve interpretability for the end users.

The work in \cite{Andrews2016} employs a visualization and animation technique for highly varied patient flows, i.e., the systematic processing of a patient from arrival to discharge at a medical facility or emergency department. The objective is to shed light on the existing differences among patient flows. To this aim, the authors propose two techniques to capture both static and dynamic behavior in a set of process variants. The static view aims at highlighting control flow differences among process variants. To this end, a process model for each process variant is discovered and, then, a configurable process model is created by merging the discovered models \cite{Rosa2009ManagingVI}. The configurable model illustrates commonalities and variant-specific paths. The dynamic view is based on animating sublogs to highlight the differences in the executions of the variants, i.e., how cases in each variant flow through the models. Similarly to the mentioned work, the paper by Conforti et al. \cite{Conforti2015} presents guidelines and a set of handy and practical examples for the analysis of process variants. Here, a configurable model is created after removing process drift behavior from the event logs to obtain a stable process behavior for each process variant.

The work by Buijs et al. in \cite{Buijs2014} proposes a technique to identify the existing deviations between process models and the corresponding executions across various organizations. This work extends the approach proposed in \cite{Buijs2011} by explicit incorporating process models in the comparative analysis. 
%The technique assumes that process models are slightly different in their structures, and it makes comparisons in two ways. In the first place, it makes 
Each process model is compared with the corresponding event log using the approach for computing alignments presented in \cite{Adriansyah.Thesis.2014}. The alignments show deviances between the models and the process executions. In addition, cross-organizational process variants are compared using an alignment-matrix where columns and rows are process models and process variants, respectively. The matrix contains the fitness values computed by aligning each process variant against the process models.

The work by van Beest et al. \cite{Beest2015} shows the behavioral distance between two sets of process executions. Behavioral differences are expressed using natural language statements highlighting exclusive frequent patterns in each set of process executions. The approach is based on encoding an event log as an annotated Event Structure \cite{Nielsen1981PetriNE}. In particular, a set of partially ordered runs (i.e., pairs of events that precede each other or are concurrent) are extracted from an event log. Each partially ordered run resembles a Prime Event Structure (PES), and the extracted set of partially runs shows causality relations. Also, a PES can be augmented with frequencies resulting in a Frequency-enhanced Prime Event Structure (FPES). The PESs of the process variants are compared by creating the Partial Synchronized Product (PSP) of the event structures \cite{Armas2014}. The PSP shows which events can be executed synchronously in two event structures identifying a mismatch if this synchronous execution is not possible.
The obtained mismatches are collected into a set of simple change patterns, which are subsequently translated into natural language statements \cite{Weber:2008}.

Cordes et al. \cite{Cordes2015} present a visualization technique that compares process variants, which is independent of a specific process modeling language. In particular, a set of process models is discovered from a set of process variants and the comparison is done over the process models. In particular, the structure of two process models is compared in a similar way as in \cite{Li2008}, i.e., by computing the minimum number of operations to transform one process model into another. The proposed algorithm compares the elements of two graphs and marks paired elements as unchanged, added, deleted, or changed to highlight the dissimilarities. Then, a view-model consistent with the input modeling language is generated for the end user. In the same vein, the work in \cite{Kriglstein2013} presents an approach independent of a specific process modeling language and based on directed graphs. The method provides some handy facilities to the end user to identify deviations. For example, the flow instance variations between two process variants can be seen in a single graph, or two process models can be compared for their structures using a difference graph model. An extension of this work is presented in \cite{Ballambettu2017}, which compares process variants using Process Maps (annotated transition systems). In a first schema, a unified Process Map is generated by considering all process variants together. A second schema generates a difference Process Map including parts that are present in one process variant but not in the others. For common elements, pair-wise differences are computed to identify parts of the Process Map that are the most peculiar of a certain process variant. 

The approach by Sun et al. \cite{Sun:2013:MER:2486046.2486067} tackles the automatic evaluation of software processes. It assumes that two process variants are available, i.e., normal and anomalous executions. Process executions are encoded into a multidimensional space. The encoding schema is similar to the unigram encoding. The main idea is to infer from the two process variants a set of contrasting itemset patterns that do not share any features. If a new process execution contains all the features of a pattern, it can be classified as normal or anomalous. Similarly, Bose et al. \cite{Bose2013} extract features such as Tandem Repeat and Maximal Repeat patterns \cite{Bose2009} to encode traces into a multidimensional vector space. Then, association rule mining and decision tree induction techniques are used to extract rules characteristic of the process execution groups. 
%It then classifies a new process execution based on the extracted rules or via classification algorithms.

Swinnen et al. in \cite{Swinnen2012} develop an approach to understand the reasons of variations in a procurement process. The proposed approach is unsupervised in the sense that process execution tags are unknown beforehand. A process model is discovered from an event log and is compared with a normative process model to uncover the differences. These differences then are used to group process executions. Then, association rule mining is used to extract rules from each group. Similarly, the work in \cite{Cuzzocrea2017DevianceDiscovery} is also unsupervised. However, this work assumes that there are two pre-defined process variants available. A model from the whole event log is discovered and is annotated with performance metrics for each process variant. Folino et al. \cite{Folino:2017:DCA:3019612.3019660} extended this work by identifying a set of rules to explain the differences between the two clusters of process executions.

The approach by Cuzzocrea et al. \cite{CuzzocreaFGP16} adopts an ensemble learning schema to find a discriminating function that classifies process executions. The strategy is to encode a single process execution into a set of vector representations, i.e., to provide a multi-view schema of each process execution. After encoding an event log in this way, a base classifier is trained for every set of vector representations. Finally, the Stacking mechanism is used to perform the classification based on the outcomes of the base classifiers. This work was extended in \cite{CuzzocreaFGP17a, Cuzzocrea2017ExtensionsAA} by identifying the label of a process execution in a probabilistic way and by extracting rules to explain the discrepancies among process variants. A follow-up work by Folino et al. \cite{FolinoEnsemble2018} proposes a peer-to-peer architecture for the discovery of base learners. The proposed architecture enables the business analyst to apply the approach in an online setting for a stream of traces. The stream of traces is processed by chunks thus allowing base learners to be adjusted periodically.

Bolt et al. \cite{Bolt2015} exploit Process Cubes \cite{BoltProcessCube2015} to split, group and compare process executions. Process cubes provide operations such as slice, dice, roll-up, and drill-down to break down process data and compare different groups or process variants to highlight dissimilarities. The work considers process executions containing the activities of a student. A process cube with various dimensions, such as ``Course code'', ``Grade'' and ``Activity Type'' is created. The outcomes of this analysis are provided in different qualitative forms such as simple statistic values, dotted charts and comparisons of activity flows. 
A follow-up of this work is presented in \cite{Bolt2016}. Here, the differences between two sets of process executions are visualized by projecting them onto a transition system where states and transitions are colored to highlight the differences. The highlighted parts only show different dominant behaviors that are statistically significant, and rare differences are masked out for the sake of readability. The transition system is annotated with information such as the frequency of an event, the elapsed time of an event (i.e., the time elapsed between the beginning of the process execution and the occurrence of the event) for each process variant. This work was extended in \cite{bolt2018} by inducting decision trees for each decision point (i.e., a node that branches) of the transition system. A set of rules is derived from the trained decision trees to explain the differences among process variants.

The work by Gulden et al. \cite{Gulden2017} proposes a circular time-line visualization, called rhythm-eye, to compare process executions in terms of execution time. In the proposed view, events are rendered as thin lines on top of the rhythm-eye ring. Average time values of each event type are represented by semi-transparent thicker circle segments, one per event type. Different event types are distinguished by colors. The approach computes a rhythm-eye view for each process variant and configures them to highlight differences.

Recently, Nguyen et al. \cite{Nguyen2018-Multiperspectquteprints117962} have proposed an approach to compare process variants via Perspective Graphs. A Perspective Graph is a graph-based abstraction of an event log where a node represents any entity referenced in an attribute of the event log (e.g., activity, resource, location), and an arc shows an arbitrary relation between entities. The approach starts by abstracting process executions in each process variant. The abstraction can be made on the order of activities or on any event attribute, e.g.\ the order in which resources hand over work to one another, or on a combination thereof (a schema). This results in a Perspective graph. The comparison can be done for any process perspective depending on the employed entities. To compare two Perspective Graphs a Differential Graph is computed. This graph contains common nodes and edges and also nodes and edges that appear in one perspective graph only. The weights of common nodes and edges are determined via statistical tests. Finally, the approach provides the identified differences in a matrix-based representation.

% The circular structure of a rhythm-eye view focuses on the normalized time interval of an activity in a process execution and abstracts from the actual length of a process. 
% A rhythm-eye view for a process cohort can be constructed by considering the average completion time of each event in that cohort. The work presents two ways for the comparison of rhythm-eye views. The first way places multiple rhythm-eye views besides each other in a common setting; the normalized projection of time intervals provided
% by the rhythm-eye views allows to compare characteristics of process cohorts. The second way nests multiple rhythm-eye views inside each other; this option creates visual patterns which allow to clearly distinguish where different sets of event data match with respect to their rhythm, and where not. The approach has been implemented as a standalone application.

\paragraph{Primary and subsumed studies}
\label{subsec: primary and subsumed}

Among the papers that successfully passed both the inclusion and exclusion criteria, we determined primary studies that constitute an original contribution to process variant analysis and deviance mining, and subsumed studies that are similar to a primary study and do not provide a substantial contribution with respect to it.
Specifically, a study is considered subsumed if:

\begin{itemize}

\item  there exists a more recent and/or more extensive version of the study from the same authors (e.g., a conference paper is subsumed by an extended journal version), or

\item it does not propose a substantial improvement/modification over a method that is documented in an earlier paper by other authors, or

\item the main contribution of the paper is a case study or a tool implementation, rather than a new method, and the method is described and/or evaluated more extensively in more recent study by other authors.

\end{itemize}

As can be seen from the \tablename~\ref{table: taxonomy}, a large number of works can considered as a primary study because of the large variety of proposed techniques.
We identified 15 primary and 14 subsumed studies.

% In the
% next section we present the primary studies in detail, and classify them using a taxonomy.

\begin{center}
\begin{table}
 \begin{tabular}{p{6cm} | p{6cm}} 
 \hline
 Primary study & Subsumed studies \\  
 \hline\hline
 Bolt et al. \cite{bolt2018} &  Bolt et al. \cite{Bolt2015, Bolt2016}  \\ 
 \hline
Ballambettu et al. \cite{Ballambettu2017}& Kriglstein et al. \cite{Kriglstein2013}  \\
 \hline
%Cordes et al. \cite{Cordes2015}  &  Li et al. \cite{Li2008}  \\
Cordes et al. \cite{Cordes2015}  &    \\
 \hline
%Poelmans et al. \cite{Poelmans2010} & Blum et al. \cite{Blum2008} \textbf{Double check}\\
 Poelmans et al. \cite{Poelmans2010} & \\
 \hline
 Partington et al. \cite{Partington:2015:PMC:2677016.2629446} & Suriadi et al. \cite{Suriadi2014, Suriadi2013}\\
 \hline
 Swinnen et al. \cite{Swinnen2012} \\
 \hline
  Sun et al. \cite{Sun:2013:MER:2486046.2486067} \\
 \hline
   Bose et al. \cite{Bose2013} \\
 \hline
Wynn et al. \cite{WYNN201793} &  Pini et al. \cite{Pini2015}, Andrews et al. \cite{Andrews2016},\newline Conforti et al. \cite{Conforti2015}\\
 \hline
 Buijs et al. \cite{Buijs2014} & Bujis et al. \cite{Buijs2011}\\
 \hline
%  van Beest et al. \cite{Beest2015} & Armas-Cervantes et al. \cite{Armas2014} \\
  van Beest et al. \cite{Beest2015} &  \\
%  \hline
%  Nguyen et al. \cite{NguyenDRMS16} & Nguyen et al. \cite{Nguyen2014}\\
 \hline
 Folino et al. \cite{FolinoEnsemble2018}  & Cuzzocrea et al. \cite{CuzzocreaFGP17a, CuzzocreaFGP16, Cuzzocrea2017DevianceDiscovery, Cuzzocrea2017ExtensionsAA}, Folino et al. \cite{ Folino:2017:DCA:3019612.3019660}\\
 \hline
 Lakshmanan et al. \cite{Lakshmanan:2013:ICC:2529737.2529772}  \\ 
  \hline
 Gulden et al. \cite{Gulden2017}  \\ 
   \hline
Nguyen et al. \cite{Nguyen2018-Multiperspectquteprints117962}  \\ 
 \hline\hline
\end{tabular}
\caption{Primary and subsumed studies}
\label{table: taxonomy}
\end{table}
\end{center}

\subsection{Input data}
\label{subsec: inputdata}
As shown in \tablename~\ref{table: summary dimension of relevant studies}, all the proposed approaches take as input an event log. 
%The timestamp attribute of the event logs not only can be used for control flow comparisons, but also to calculate the completion time of activities and cases. 
%in a case, or the cycle time of a case. This information can then be used to extend the corresponding analysis. 
The input event log may have a prior structure that can be used to identify process variants, or process variants can be created based on event attributes such as resources (see \figurename~\ref{fig:general framework}). Some approaches also require a process model as input. In the following, we explain how the selected works employ input data in their analysis.

Some works assume that process executions are grouped or tagged beforehand. For example, Sun et al. \cite{Sun:2013:MER:2486046.2486067} take as input two sets of software process executions, i.e., normal and anomalous executions. Suriadi et al. \cite{Suriadi2014} use four groups of process executions coming from four different hospitals. Similarly, the process variants in \cite{Buijs2014, Buijs2011, Beest2015, Andrews2016, CuzzocreaFGP16, CuzzocreaFGP17a, Cuzzocrea2017ExtensionsAA, FolinoEnsemble2018, bolt2018, Gulden2017, Ballambettu2017} are pre-defined. Although the input process executions in \cite{Nguyen2018-Multiperspectquteprints117962} are grouped beforehand, the approach can inherently create process variants based on performance attributes.

In other studies, process variants can be created based on performance data. The studies in \cite{Pini2015, WYNN201793} use min/max/avg activity durations as objective measures to characterize different process variants. Cordes et al. \cite{Cordes2015}, in their analysis, employ case attributes, such as the age or the region of a customer, to group together process executions. Likewise, Suriadi at al. \cite{Suriadi2013} use the cycle time of a case to group process executions into cohorts. Bose et al. \cite{Bose2013} group process executions of a process to repair malfunctions in X-ray machines according to the mean-time-to-repair of the parts that must be replaced. The work by Bolt et al. \cite{Bolt2015} uses Process Cubes \cite{BoltProcessCube2015} to group process executions based on performance data of students. 
%In that study, a student's set of activities based on ``Course code'' or ``Grade'' is considered as a process execution. 

The works in \cite{Folino:2017:DCA:3019612.3019660, Swinnen2012, Cuzzocrea2017DevianceDiscovery} neither take as input a categorized set of process executions nor group them based on event or case attribute values. Indeed, the main aim of such studies is to discover process variants with no prior knowledge. However, the study in \cite{Cuzzocrea2017DevianceDiscovery} assumes as prior knowledge the percentage of deviant and non-deviant cases.

Some approaches take as additional input a normative process model \cite{Kriglstein2013, Buijs2014, WYNN201793}. A normative process model is used as a reference model for quantifying to what extent the process variants differ from normative executions. The normative process model can be provided using different notations. The authors in \cite{Buijs2014} use BPMN, whereas the authors in \cite{WYNN201793} employ Petri nets. The approach presented in \cite{Kriglstein2013} does not pose any specific restrictions on the process modeling language employed, but for special concepts of certain languages developing extensions could become necessary.

\subsection{Outcomes}
\label{subseq:output}
The outputs of process variant analysis depend on the research questions and objectives considered in the different studies, and vary across different domains. However, as shown in \tablename~\ref{table: summary dimension of relevant studies}, most of the works focus on providing explainable results showing how process variants differ from different perspectives. In particular, the outcomes of process variant analysis can be grouped based on the following categories:

\begin{itemize}
    \item \textbf{Rule-based}: The works in \cite{Swinnen2012, Suriadi2013, Sun:2013:MER:2486046.2486067, Bose2013, Folino:2017:DCA:3019612.3019660} represent the existing discrepancies among process variants through a set of rules or causal relations. All these works provide the extracted rules according to different encoding schemas, but always as a conjunction of a set of antecedents, and a consequent that discriminates among different process variants. Similarly, the work in \cite{Poelmans2010} finds itemsets, i.e., sets of activities, that differ for different process variants, whereas, the work in \cite{Lakshmanan:2013:ICC:2529737.2529772} finds frequent patterns characteristic of each process variant. Also, the analysis in \cite{bolt2018} extracts a set of rules that can be used to assign a cohort label to each process execution. van Beest et al. \cite{Beest2015} generate discriminative rules in terms of natural language statements.
    
    \item \textbf{Model-based}: A significant number of works provide as outputs process models that are easy to interpret for end users. The output is either a set of process models representing the behavior of the different process variants or an embodiment process model representing the behavior of an entire event log. The discovered process models are usually annotated with performance data. For example, the works in \cite{Suriadi2013, Bolt2015, Bolt2016, bolt2018} annotate the discovered transition systems with the frequency of the transitions between two states, whereas the approaches in \cite{Cuzzocrea2017DevianceDiscovery, Folino:2017:DCA:3019612.3019660} annotate the discovered transition systems with performance data such as elapsed or remaining time. Andrews et al. \cite{Andrews2016} annotate the discovered configurable BPMN model with the \emph{length of stay} of a patient in a hospital. The work in \cite{Partington:2015:PMC:2677016.2629446} generates BPMN models annotated with performance data and a transition system for each process variant highlighting frequent paths. In the same way, Lakshmanan et al. \cite{Lakshmanan:2013:ICC:2529737.2529772} superimpose frequent patterns on the discovered transition systems. Suriadi et al. \cite{Suriadi2014} derive Petri nets from the discovered transition systems representing the behavior of the process variants. Ballambettu et al. \cite{Ballambettu2017} use annotated transition systems to highlight differences among process variants. The approaches by Kriglstein et al. \cite{Kriglstein2013} and Cordes et al. \cite{Cordes2015} are more flexible and generate an annotated directed graph that can be translated into other notations. The study in \cite{Kriglstein2013} can also be provided with an input process model, which is annotated with the differences among the process variants. Similarly, Pini et al. \cite{Pini2015} annotate an input transition system with various performance data such as the median execution times of activities. This work was extended by Wynn et al.\ in \cite{WYNN201793} where the input Petri net is projected into a flat model annotated with performance data such as waiting time between activities.  
    
    \item \textbf{Descriptive}: Some works provide visual summaries and descriptive statistics for performance data or event attributes to highlight the differences among process variants. These outputs are standalone or can be integrated with the other outcomes, e.g., they can be used to annotate process models as mentioned earlier. Examples of techniques that use standalone descriptive statistics are \cite{Suriadi2013} where basics statistics are used to identify which cases are more complex than others and Bolt et al. \cite{Bolt2015} that employ bar charts to show the number of students in each process variant and use dotted charts to visualize how many videos are watched by students in different cohorts. Poelmans et al. \cite{Poelmans2010} identify process variants based on the patient's \emph{length of stay} in a hospital and then tabulate some important factors such as the number of cases and the average number of activities per case in each process variant. Also, several studies compare the control flow characteristics of process variants in tabular form \cite{Buijs2011, Buijs2014, Suriadi2014, WYNN201793, Pini2015}. The table includes fitness values showing how well a process execution from one cohort can be replayed by representative models of other process variants. Nguyen at al. \cite{Nguyen2018-Multiperspectquteprints117962} use a matrix-based structure for displaying statistically significant discrepancies among process variants derived from a Differential Graph. Gulden et al. \cite{Gulden2017} provide a circular visualization, called rhythm-eye, to compare the control flow structures of different process variants.

\end{itemize}

Some recent works provide a labeling or classifications of process executions. In these works, the outcome is a class label \cite{CuzzocreaFGP16}, or a set of probabilities that show how a process execution associates to different groups \cite{CuzzocreaFGP17a, Cuzzocrea2017ExtensionsAA, FolinoEnsemble2018}. It is worth mentioning that the aim of these works is training a classifier for each process variant to label upcoming completed process executions. This approach is different from  predictive process monitoring techniques, which predict the outcome of an ongoing process execution or estimate the required time to complete. Indeed, predictive process monitoring techniques operate in an online setting, whereas the mentioned studies operate in an offline setup.

\subsection{Type of analysis}
\label{subsec:Bariants analsysis}
To conduct process variant analysis of process executions, different perspectives of the process under analysis can be taken into consideration. The process perspectives to look at in variant analysis depend on factors such as the research questions addressed and the availability of data. These perspectives also determine the type of outcome that needs to be produced and the underlying algorithms that need to be developed.
%that can deal with the specific perspectives that need to be investigated.

%Nonetheless, this sort of analysis is a bit overshadowed in between. Thus, to stress the importance of variants analysis, this part is designated to explore and to provide more insight about this issue.

Based on the perspectives investigated, we can classify the types of analysis as: 

\begin{itemize}
    \item \textbf{Control flow}: In this type of analysis, a process execution is considered as an ordered set of activities discarding all available related contextual or performance attributes. Some of the studies that use the control flow perspective \cite{Poelmans2010, Swinnen2012, Suriadi2013, Sun:2013:MER:2486046.2486067, Bose2013, Lakshmanan:2013:ICC:2529737.2529772, Beest2015,Folino:2017:DCA:3019612.3019660,bolt2018} generate a set of rules or patterns to express the control flow discrepancies in a set of process executions. Other works \cite{Kriglstein2013,Suriadi2014,Partington:2015:PMC:2677016.2629446,Cordes2015,Pini2015,Bolt2015,Bolt2016,Andrews2016, Cuzzocrea2017DevianceDiscovery,WYNN201793, Ballambettu2017} extract process models from logs representing the behaviors of different process variants. Some works provide a visual comparison to highlight discrepancies. For example, the work in \cite{Nguyen2018-Multiperspectquteprints117962} provides a compact matrix-based representation of statistically significant differences from a Differential Graph. Similarly,
    Gulden et al. \cite{Gulden2017} produces rhythm-eye views to compare process variants based on control flow.
    Finally, several studies \cite{Buijs2011, Buijs2014, Suriadi2014, WYNN201793, Pini2015} compare the control flow characteristics of process variants using alignments. 
    
    \item  \textbf{Performance analysis}: Recent works have focused more on the analysis of contextual or performance attributes. This perspective is important since a set of process executions with the same control flow could have different cycle times or use different types of resources. Most of these works consider time-related performance data in their analysis. For example, Poelmans et al. \cite{Poelmans2010} consider the \emph{length of stay} of a patient for cycle time analysis to discover discrepancies among patients with the same control flow structures. In \cite{Suriadi2014,Partington:2015:PMC:2677016.2629446,Pini2015,Andrews2016,Cuzzocrea2017DevianceDiscovery,Folino:2017:DCA:3019612.3019660}, the authors take into account the cycle time of process executions to separate process executions into groups and then find control flow characteristics of slow cases. 
    %Likewise, Folino et al. \cite{Folino:2017:DCA:3019612.3019660} use performance data such as cycle time of a case and workloads to infer a set of rules that describe different control flow models. 
    In the same vein, the work by Nguyen et al. \cite{Nguyen2018-Multiperspectquteprints117962} discovers a control flow model for any combination of time-based attribute values. The studies in \cite{Suriadi2014,Partington:2015:PMC:2677016.2629446,WYNN201793,Gulden2017, Ballambettu2017} work with the waiting times between activities across different process variants to understand the existing performance variations, whereas Pini et al. \cite{Pini2015} consider the median duration of each activity. Bolt et al. \cite{Bolt2016, bolt2018} investigate the elapsed time, i.e., the time between the starting point of a process execution and the occurrence of a certain event to identify performance deviations. 
    Other works start from pre-defined groups of process executions and leverage both control flow and performance data to characterize those groups. For example, in \cite{CuzzocreaFGP16,CuzzocreaFGP17a,Cuzzocrea2017ExtensionsAA,FolinoEnsemble2018}, the authors use both control flow and cycle time of process executions to train an ensemble classifier. The classifier assigns an upcoming process execution to a process variant.
\end{itemize}

Except the work in \cite{Nguyen2018-Multiperspectquteprints117962}, it is interesting to observe that none of the techniques we retrieved considers the possibility of comparing process variants along other perspectives besides the above two. Yet, it is conceivable that two process variants may differ along the resource perspective (e.g. different resource pool), or along the data perspective.

\subsection{Family of algorithms}
\label{subsect: family of algorithms}
When conducting process variant analysis, the underlying algorithms used are strongly influenced by the input data and by the accessibility of performance attributes. Nevertheless, the proposed algorithms share the ability of providing explainable results. Broadly speaking, the algorithms used in the selected papers belong to two main families: 

\begin{itemize}
       \item \textbf{Process mining}: This family of algorithms uses process mining techniques to uncover differences among process variants. The majority of the proposed approaches discover a process model for every process variant, and then compare them to highlight the differences. %, or compare the discovered process models with the expected or normative behavior described in an input model. 
       In \cite{Suriadi2013, Bolt2015,Bolt2016}, the authors discover an annotated transition system where states and transitions are colored to show different dominant behaviors (representing different process variants) that are statistically significant. Factors such as frequency and elapsed time of an event are considered in the analysis. Similarly, in \cite{Cuzzocrea2017DevianceDiscovery}, a transition system is discovered from the whole event log and annotated with performance metrics characterizing each process variant. Kriglstein et al. \cite{Kriglstein2013} compute a directed graph, called Difference model, to highlight the existing differences between two process variants. In this work, a normative process model representing the expected process behavior can be provided as input. This work was extended by Ballambettu et al. \cite{Ballambettu2017} where annotated transition systems are used (called process maps) to represent the behaviors of different process variants. Suriadi et al. \cite{Suriadi2014} discover a Petri net for every process variant and quantify the closeness of their control flow structures using alignments \cite{Adriansyah.Thesis.2014}. In particular, alignments provide a fitness value that tells how good a process model discovered for a process variant can replay the observed executions available in the logs corresponding to the other cohorts. In the same way, the works in \cite{Buijs2011, Buijs2014, Pini2015, WYNN201793} compute alignments to come across the existing control flow differences among process variants. Most of these works take as input a normative model representing the expected behavior of the process. The analysis by Partington et al. \cite{Partington:2015:PMC:2677016.2629446} first discovers a process model using the Fuzzy Miner from the whole event log, then it replays process executions of different process variants on the discovered model to characterize them using infrequent-traversed paths. Cordes et al. \cite{Cordes2015} compare the structures of two process models discovered from two process variants using TGraphs \cite{Ebert95}. A TGraph is an intermediate representation of a process model, wherein no distinction among different types of nodes and different types of edges is assumed. Each node and edge, however, carries additional information to preserve the semantics of the original process model. For example, in a Petri net, a node can be marked as a transition or a place. Two TGraphs are compared using the Snapshot-diff algorithm \cite{Labio96efficientsnapshot}, which produces a Difference model. In particular, the algorithm compares two graphs by comparing their elements and marking them as unchanged, added, deleted, or changed to highlight dissimilarities. van Beest et al. \cite{Beest2015} encode an event log as an \textit{annotated event structure} \cite{Nielsen1981PetriNE}, which is a directed acyclic graph where nodes represent event occurrences sharing a common history. Annotated event structures also keep information about the frequency of each event. The technique extracts a set of partially ordered runs where pairs of events can precede each other or be concurrent. Each partially ordered run resembles a \textit{prime event structure}, i.e., a graph of events representing the causal relations between events. The partially ordered runs are merged to derive a prime event structure of the full log. When different logs corresponding to different process variants are available different prime event structures are derived using the above procedure and then compared using the \textit{partial synchronized product} of the event structures \cite{Armas2014}. The identified mismatches are collected into a set of simple change patterns, which are subsequently translated into natural language statements \cite{Weber:2008}. Andrews et al. \cite{Andrews2016} discover a BPMN process model for each process variant, and then build a configurable process model obtained by merging the discovered models using the technique proposed in \cite{Rosa2009ManagingVI}. The configurable model illustrates commonalities and variant-specific paths. The paper also proposes a log replaying technique using a heuristic-based backtracking algorithm to compare a process execution and a BPMN model. Nguyen et al. \cite{Nguyen2018-Multiperspectquteprints117962} discover, from the process executions corresponding to a process variant, a perspective graph taking into consideration control flow and different combinations of performance attributes. Then, two perspective graphs are compared and merged into a Differential Graph in which the elements that are statistically different in the perspective graphs are highlighted.
    Finally, Poelmans et al. \cite{Poelmans2010} use the \textit{formal concept analysis} \cite{Ganter:1997:FCA:550737} to capture a representative set of activities in each process variant. Formal concept analysis is a method for deriving implicit relationships between objects (in process variant analysis activities) described through a set of attributes. 
     \item \textbf{Machine learning}: This family of algorithms exploits machine learning or statistical algorithms to analyze process variants. 
    Sun et al. \cite{Sun:2013:MER:2486046.2486067} use contrast itemsets \cite{Bay2001ContrastMining} to characterize process variants. Contrast itemsets are composed of attribute values that differ across groups of process executions. 
    Bose et al. \cite{Bose2013} transform process executions into multidimensional vector representations using as features frequent control flow patterns. Then, they apply association rule mining and decision tree induction to infer a set of rules that characterize process variants. Lakshmanan et al. \cite{Lakshmanan:2013:ICC:2529737.2529772}  find frequent sequence patterns using Sequential Pattern Mining with bitmap representation (SPAM) \cite{Ayres(SPAM)2002}. The patterns are used to represent every process execution as a Bag-of-Pattern (BoP). Then, Density-Based Spatial Clustering of Applications with Noise (DBSCAN) \cite{DBSCAN} is used to cluster process executions in different cohorts. The work in \cite{CuzzocreaFGP16}, after transforming process executions into multidimensional feature vectors, adopts an ensemble method (Bayesian Model Averaging) to learn a classifier via stacking \cite{Dietterich:2000:EMM:648054.743935}. Stacking is a meta-learning task in machine learning where a classifier uses the output of other base classifiers to better classify or label a process execution. In particular, meta-learning allows a learner to not only learn from historical data, but also from other learning tasks. The approaches in \cite{CuzzocreaFGP17a, Cuzzocrea2017ExtensionsAA} extend the previous work by adopting the Hidden Naive Bayes classifier \cite{Zhang:2005:HNB:1619410.1619480} at the meta-learning level. This type of classifier provides probabilistic outcomes. Folino et al. \cite{FolinoEnsemble2018} extend the previous works by proposing a peer-to-peer computing architecture to speed up the training phase of base learners.
\end{itemize}

It is worth pointing out that, though we broke up the process variant analysis algorithms into two families, some works belong to both. For example, Swinnen et al. \cite{Swinnen2012} first discover a process model using the Fuzzy Miner, and then find discrepancies between the discovered model and a normative model to assign process executions to different process variants. Then, the authors use the Apriori algorithm \cite{PredictiveApriori} from association rule mining to find a set of rules characterizing each process variant. Similarly, Folino et al. \cite{Folino:2017:DCA:3019612.3019660} propose an iterative optimization algorithm to infer a set of rules to group process executions into process variants. Then, a process model is discovered for each cohort using the Fuzzy Miner. Works that are in between the two families are the one presented in \cite{Suriadi2013} that infers a set of causal relation rules to characterize lengthy process executions and the analysis presented in \cite{Suriadi2014} that uses K-means clustering to group the input set of process executions. Finally, Bolt et al. \cite{bolt2018} also use a typical process mining algorithm to create an annotated transition system starting from a log, and then, for every decision point in the transition system, train a classifier to distinguish different process variants.

\subsection{Evaluation data and application domain}
\label{Subsec:evaluation and domain}
As reported in \tablename~\ref{table: summary dimension of relevant studies}, most of the surveyed methods have been validated on at least one real-life event log, and a few studies were additionally validated on simulated (synthetic) logs. Most of the real-life logs employed are publicly available in the 4TU Center for Research Data\footnote{\url{https://data.4tu.nl/repository/collection:event_logs_real}}. Among the methods that use real-life logs, we observed a growing trend to use publicly available datasets, as opposed to private logs that hinder the reproducibility of the results.

Process variant analysis is attractive and beneficial in domains where a single process model is executed across different organizations. A good example is provided by SaaS applications, where a single version of an application, with a single configuration, is used for different customers, such as applications for logistics, Incidence Management (IcM), financial management and healthcare management. From \tablename~\ref{table: summary dimension of relevant studies}, we notice that most of the selected works pertain to healthcare (12 studies), logistics (4 studies), public administration (5 studies), industrial and insurance organizations (5 studies), financial institutions (3 studies), education systems (1 study) and IcM systems (1 study).

\subsection{Implementation}
\label{Subsec: implementation}

Providing publicly available implementations and experimental data facilitates the reproducibility of the results and enables researchers to build on past works. According to \tablename~\ref{table: summary dimension of relevant studies}, around half of the methods provide an implementation as a plug-in of the process mining tools ProM \cite{ProMFramework} and Apromore \cite{Apromore}. Both the aforementioned frameworks are open-source and portable, which allows researchers to easily develop and test new algorithms. Similarly, a few works employ other tools such as Disco\footnote{\url{https://fluxicon.com/disco/}}, Nitro\footnote{\url{https://fluxicon.com/nitro/}}, and RapidProM\footnote{\url{http://rapidprom.org/}}.

Several techniques that employ machine learning algorithms use Weka \cite{WEKA}, which is an open-source library implementing machine learning algorithms. Other machine-learning-based approaches use the Hidden Markov Model toolbox for Matlab\footnote{\url{https://www.cs.ubc.ca/~murphyk/Software/HMM/hmm.html}} and RapidMiner \cite{Rapidminer}.

Finally, some works implemented their methods as standalone applications; others did not provide any prototype at all, or provided only parts of them.

\section{UNIFYING FRAMEWORK}
\label{sec: methodology framework}
As outlined in the previous section, a wide range of methods have been proposed to tackle the problem of process variant analysis. However, because of the heterogeneous nature of the underlying algorithms, their inputs, and their outputs, the classification proposed in the previous section, while comprehensive, does not provide us with a unifying view of the state of the art in the field.

As a first step towards building a unifying view of the field, we propose an alternative classification of existing methods based on the observation that some of the methods seek to identify discriminating characteristics or patterns, while other approaches discover a model of each of the variants and then compare the variants based on the discovered models. This observation leads us to classify existing approaches into three categories: \textit{discriminative, generative} and \textit{hybrid}. This broad classification is a step towards unifying the various strands of research in the field, by bringing them together in terms of their underpinning paradigms. Below, we provide a detailed explanation of each of these three categories.

\begin{table}[t!]
\footnotesize{\begin{tabular}{|l|l|l|l|l|l|l|l|l|l|}
\hline
Case\_id & City & Sex & Product & Cycle time (s) & Order & Pay in cash & Pay by card & Approval & Disapproval \\ \hline
1        & NY   & M   & Book    & 183        & 1     & 1           & 0           & 1        & 0           \\ \hline
2        & MA   & F   & Sofa    & 58,690     & 1     & 1           & 1           & 1        & 1           \\ \hline
3        & LA   & M   & T.V.    & 960        & 1     & 0           & 1           & 1        & 0           \\ \hline
4        & LA   & F   & Book    & 960        & 1     & 1           & 0           & 1        & 0           \\ \hline
\end{tabular}}
\caption{Encoding process executions using unigram}
\label{table:unigram}
\end{table}

\begin{table}[t!]
\tiny{\begin{tabular}{|l|l|l|l|l|l|l|l|l|l|l|l|}
\hline
Case\_id & City & Sex & Product & Cycle time & \begin{tabular}[c]{@{}l@{}}Order \\ Pay in cash\end{tabular} & \begin{tabular}[c]{@{}l@{}}Order\\ Approval\end{tabular} & \begin{tabular}[c]{@{}l@{}}Order\\ Pay by card\end{tabular} & \begin{tabular}[c]{@{}l@{}}Pay in cash\\ Approval\end{tabular} & \begin{tabular}[c]{@{}l@{}}Disapproval \\ Pay in cash\end{tabular} & \begin{tabular}[c]{@{}l@{}}Pay by card\\ Disapproval\end{tabular} & \begin{tabular}[c]{@{}l@{}}Approval \\ Pay by card\end{tabular} \\ \hline
1        & NY   & M   & Book    & 183        & 1                                                            & 1                                                        & 0                                                           & 0                                                              & 0                                                                  & 0                                                                 & 0                                                               \\ \hline
2        & MA   & F   & Sofa    & 58,690     & 0                                                            & 0                                                        & 1                                                           & 1                                                              & 1                                                                  & 1                                                                 & 0                                                               \\ \hline
3        & LA   & M   & T.V.    & 960        & 0                                                            & 1                                                        & 0                                                           & 0                                                              & 0                                                                  & 0                                                                 & 1                                                               \\ \hline
4        & LA   & F   & Book    & 960        & 1                                                            & 0                                                        & 0                                                           & 1                                                              & 0                                                                  & 0                                                                 & 0                                                               \\ \hline
\end{tabular}}
\caption{Encoding process executions using bigram}
\label{table:bigram}
\end{table}

\subsection{Discriminative}
\label{subsec: discriminative}

A discriminative approach to process variant analysis leverages techniques that aim at identifying features or patterns that can be extracted from process executions directly to discriminate among process variants and highlights the existing differences. These features include both control flow features and performance attributes and can range from frequency of individual activities/attributes, itemsets of activities/attributes, prefixes of process executions or their subsequences or, a combination of them.

In general, these approaches can use two mechanisms to infer discriminatory features:

\begin{itemize}
    \item \textbf{Vector-based}: 
This mechanism encodes every process execution into a vector representation labeled either with the corresponding process variant (to discriminate among different cohorts) or with a performance attribute (to discriminate among different values of this attribute within the same cohort). Then, a classifier is trained using these multidimensional representations of process executions. The trained model aims at identifying which dimension or combination of dimensions of the input vectors better contribute to the determination of the label. The crucial part of this mechanism is that a process execution is encoded into a vector representation, i.e., $g: \sigma \rightarrow \mathbf{x}$. There are several techniques that use this mechanism, though most of them use \textit{lossy encodings}. A lossy encoding does not capture the entire information of a process execution when it is transformed into a feature vector, thus some information can be lost during the transformation. One easy way for implementing a lossy encoding is by using \textit{n-grams}. An n-gram is a sequence of $n$ items. For example, for the event log presented in Table~\ref{table:event log example}, the corresponding unigram and bigram representations alongside with performance attributes are presented in Tables~\ref{table:unigram} and \ref{table:bigram}. It is easy to see why an n-gram is a lossy encoding. Indeed, the unigram encoding in \tablename~\ref{table:unigram}  ignores the existing order of activities in the process executions, i.e., it considers a process execution as a bag of activities. It is clear that the n-gram encoding for $n \geq 2$ better captures the activity orders although it increases the \textit{curse of dimensionality}. Folino et al. \cite{FolinoEnsemble2018} employ n-gram encodings for $n\in [1,4]$ to examine which patterns better contribute to the prediction of cohorts. It is noteworthy that there are several more sophisticated encoding schemas that better capture the behavior observed in process executions. For example, Sun et al. \cite{Sun:2013:MER:2486046.2486067} use a modified version of unigrams where the position of an element is also considered as a dimension, whereas Bose et al. \cite{Bose2013} use tandem repeats and maximal repeats patterns to encode traces into multidimensional vectors. In their works, Cuzzocrea et al. \cite{CuzzocreaFGP16, CuzzocreaFGP17a, Cuzzocrea2017ExtensionsAA} consider different combinations of these patterns. Similarly, Nguyen et al. \cite{NguyenDRMS16} extensively experimented different encoding schemas using different combinations of features. The classifiers used to classify feature vectors (see \figurename~\ref{fig:accuracy-explainability}) range from rule-based classifiers (high explainability, low accuracy) to ensemble learning algorithms (low explainability, high accuracy).

\begin{figure}[t!]
	\centering
	\includegraphics[width=1\linewidth]{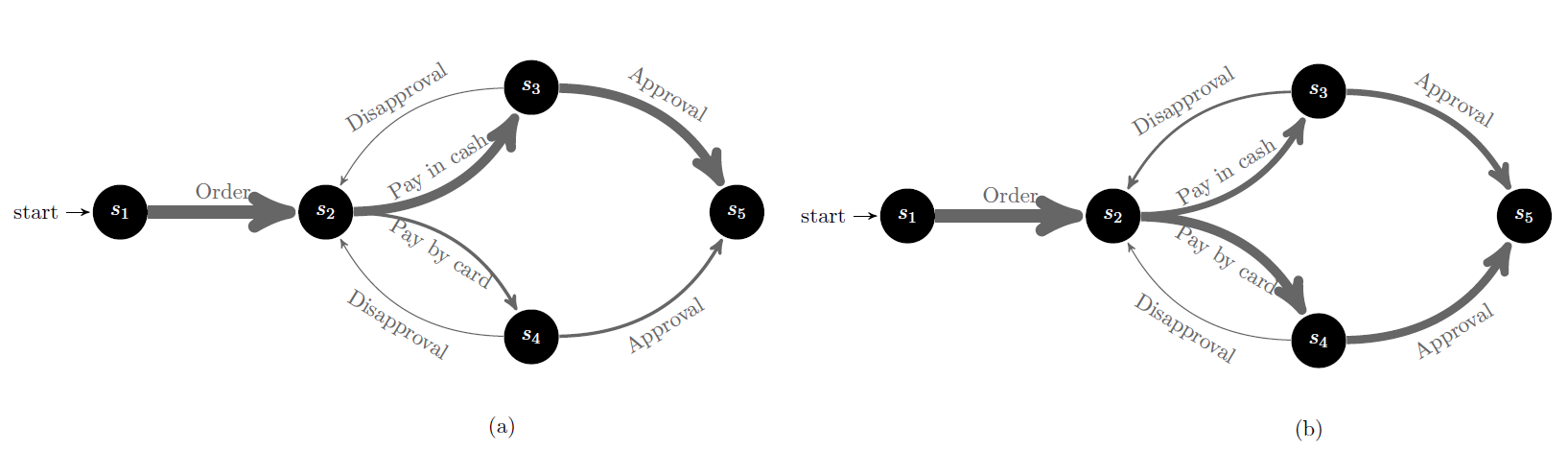}
	\caption{Replaying two different process variants on the input process model }
	\label{fig:replay example}
\end{figure}

\item \textbf{Model-based}: This mechanism uses an input process model, and considers it as the normative behavior. The main idea is to determine whether the observed behavior in each process variant, i.e., a process execution, agrees with the expected behavior or not. To implement this mechanism, two similar techniques can be used, namely alignment analysis and log replay. Although computing alignments is optimal in finding deviations, its complexity is exponential \cite{Adriansyah.Thesis.2014}. Therefore, log replaying methods, having a lower complexity, can be leveraged to identify deviations. Using log replay, it is possible to monitor the frequency of every process path observed in an event log. Thus, frequent and infrequent paths can be determined and used to highlight the discrepancies among the behaviors of process variants and between the behavior of each process variant and the normative model. Figure \ref{fig:replay example} shows the frequencies of paths on an input process model after replaying the process executions of two different process variants on the model. The thickness of an edge shows how many times the edge has been traversed by process executions in a process variant so that it is easy to extract frequent paths characteristic of each process variant. \end{itemize}

\begin{figure}[t!]
	\centering
	\includegraphics[width=1\linewidth]{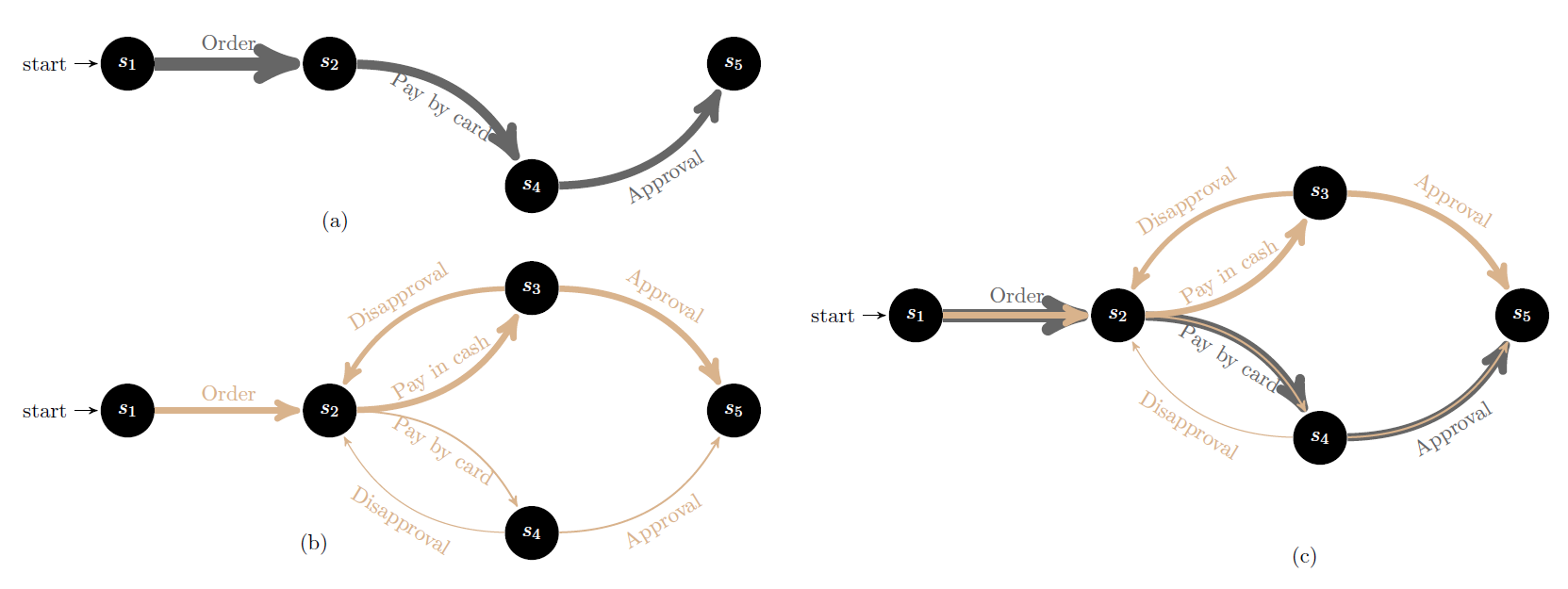}
	\caption{Merging two discovered process models into one}
	\label{fig:merging example}
\end{figure}

\subsection{Generative}
\label{subsec: generative}
A generative approach to process variant analysis leverages process model comparison techniques to shed light on existing differences among process variants. These approaches, usually, do not represent discrepancies in terms of patterns or rules as in the descriptive approach; instead, they present discrepancies graphically.

In general, a generative approach is composed of two stages. In the first stage, a process model for every process variant is discovered. The discovered model can be represented using different formalisms such as Petri nets, transition systems, BPMN models, Hidden Markov Models. In the second stage, the discovered models from each process variant are compared with each other or with a normative process model. In most of the cases, the discovered process models are merged into a single model where the behaviors of the single cohorts are highlighted \cite{Cordes2015, Kriglstein2013, Andrews2016, bolt2018, Bolt2016}. 

There are several sophisticated methods for merging process models that are beneficial for generative approaches \cite{Rosa2009ManagingVI}. However, a straightforward way for merging two process models representing two process variants was presented in \cite{Kriglstein2013, Ballambettu2017} and is illustrated in Figures \ref{fig:merging example} (a), (b) and (c). Figures \ref{fig:merging example} (a), (b) show the control flow structures and the corresponding path frequencies of two different process variants. To have the representation of both behaviors in a single process model, the two models can be merged as shown in \figurename~\ref{fig:merging example}(c). 

The main advantage of generative approaches over discriminative approaches is not only related to the readability of the results that are easier to understand for end users, but also to their lower sensitiveness to noise. This is due to the fact that the process discovery techniques used in these approaches can be seen as a filtering or pre-processing step that pull out noise or unusual behaviors before identifying the discrepancies, which leads to having more comprehensible results.

\subsection{Hybrid}
\label{subsec: hybrid}
Hybrid approaches are a combination of a generative phase and a discriminative phase. The idea behind hybrid approaches is to discover discriminatory patterns or rules and project them onto a process model. These approaches are usually composed of several stages. Usually a hybrid approach starts by discovering a process model from the log corresponding to a process variant and, then, discriminative patterns are discovered to characterize different process variants. Finally, the discriminative patterns are superimposed on the discovered model to highlight discriminative parts. For example, the approach presented in \cite{Lakshmanan:2013:ICC:2529737.2529772} finds frequent sequence patterns using Sequential Pattern Mining and project them onto the process model discovered from the process executions of each process variant.

One of the most straightforward hybrid approaches consists in discovering a process model for every process variant and then applying cross-validation. Usually, cross-validation is accomplished by computing alignments to quantify how similar control flow structures are across different process variants \cite{Suriadi2014, Pini2015, Buijs2014}. For example, assume that there are $m$ process variants, and $m$ process models representing them are discovered. Then, a process variant is selected, and its process executions are aligned with the other $m-1$ process models. The procedure repeats for all process variants. The result is a matrix structure containing the average fitness values, which show the similarities in terms of control flow of the different process variants.

\section{CONCLUSION}
\label{sec: conclusion}

Understanding the differences between multiple process variants can help analysts and managers to make informed decisions as to how to standardize or otherwise improve a business process, for example by helping them find out what factors lead to a given variant exhibiting better performance than another one. Various methods for process variant analysis based on event logs have been proposed in the past decade. However, to this date, the field remains rather fragmented.

%Examining the deviations that might exist among the executions of a process is the subject of business process variant analysis. This field has various applications in diverse domains, from Cloud Computing where different platforms execute the same process for different customers, to compliance checking of Service-Level Agreement violations. Manifesting such discrepancies among process executions not only can prevent extra costs or even permanent loss but, also provides an infrastructure for adopting policies for the evolution of a process.

As a first step towards building up a unified view of the field, this article provided a survey and a classification of existing methods for business process variant analysis. The relevant studies were identified through a systematic literature review, which retrieved 29 studies. Out of these 29 studies, 15 of them propose distinct methods (primary studies). Through  further analysis of the primary studies, a taxonomy was proposed based on four aspects: (1) the type of input data required; (2) the provided outputs; (3) the type of analysis, and (4) the algorithms employed. While analyzing the algorithms employed, we noticed that some of the methods rely on the identification of characteristics or patterns that are frequently present in one variant and not in the other variants (discriminative approaches). Other approaches, in contrast, seek to discover a model for each of the process variants and then compare the discovered models (generative approaches). It was found that 8 out of the 15 primary studies employ a generative approach, while the remaining 9 employ discriminative or hybrid generative-discriminative approaches.

The study shed light into research gaps in the field and corresponding avenues for future work. First, most of the studies consider time-related performance, thus ignoring other performance dimensions such as cost, quality, flexibility, or compliance. Second, while a large subset of existing approaches focus on control-flow differences, the question of comparing process variants along the data perspective or the resource perspective has not received attention. Finally, most of the proposed approaches show the identified deviations in a descriptive way, without backing up the detected differences between process variants with statistical tests or causal analysis, which could help to generate recommendations for addressing deficiencies in one or more of the analyzed process variants. In other words, this study calls for the development of multi-perspective approaches to process variant analysis, which would seek not only to identify differences between two or more variants, but also, to conclusively determine which of these differences contribute to observed differences in the performance of the process variants. Such multi-perspective and statistically grounded approaches could help analysts and managers to obtain insights into how to improve the performance of specific variants of a business process.

\begin{acks}
This research is partly funded by the Australian Research Council (grant DP180102839) and the European Research Council (PIX project).
\end{acks}

%
% The next two lines define the bibliography style to be used, and the bibliography file.
\bibliographystyle{ACM-Reference-Format}
\bibliography{sample-base}

% 
% If your work has an appendix, this is the place to put it.
\appendix

% \section{Research Methods}

% \subsection{Part One}

% Lorem ipsum dolor sit amet, consectetur adipiscing elit. Morbi malesuada, quam in pulvinar varius, metus nunc fermentum urna, id sollicitudin purus odio sit amet enim. Aliquam ullamcorper eu ipsum vel mollis. Curabitur quis dictum nisl. Phasellus vel semper risus, et lacinia dolor. Integer ultricies commodo sem nec semper. 

\end{document}